{}%disable hyper at arxiv
{}%use pdflatex at arxiv

\documentclass[12pt,a4paper]{article}

\newif\ifpublic\publicfalse
\newif\ifniklas\niklastrue
\newif\ifniklas\niklastrue
\usepackage{ifpdf}
\ifniklas\ifpdf\publictrue\else\publicfalse
\PassOptionsToPackage{hypertex}{hyperref}
\PassOptionsToPackage{draft}{graphicx}
\fi\fi

\usepackage{subfigure}
\usepackage{color}
\usepackage{tikz}
\usetikzlibrary{decorations.markings}
\usetikzlibrary{shapes,arrows,automata,positioning}
\usepackage{tikz-3dplot}

\tikzstyle{decision} = [diamond, draw, fill=blue!20, 
    text width=4.5em, text badly centered, node distance=3cm, inner sep=0pt]
\tikzstyle{block} = [rectangle, draw, fill=blue!20, 
    text width=5em, text centered, rounded corners, minimum height=4em]
\tikzstyle{line} = [draw, -latex']
\tikzstyle{cloud} = [draw, ellipse,fill=red!20, node distance=3cm,
    minimum height=2em]

\usepackage[utf8]{inputenc}
 \usepackage[T1]{fontenc}
\usepackage{amsmath,amssymb,amsthm}
\ifpublic\else\usepackage{showkeys}\fi
\usepackage[bookmarks=true,hyperfigures=true]{hyperref}
\usepackage{graphicx}
\usepackage{dsfont}
\usepackage{xcolor}
\usepackage[bulletsep]{collref}
\usepackage{bm}
\usepackage{multirow}
\usepackage{enumitem}
\usepackage[numbers]{natbib}
\ifniklas

\def\showkeysrefformat#1{{\normalfont\tiny\ttfamily#1}}
\makeatletter
\def\SK@@ref#1>#2\SK@{%
 {\@inlabelfalse\leavevmode\vbox to\z@{%
 \vss\SK@refcolor\rlap{\vrule\raise .75em%
  \hbox{\showkeysrefformat{#2}}}}}}
\makeatother
\fi

\usepackage[a4paper,text={160mm,247mm},centering]{geometry}
\allowdisplaybreaks[3]

\numberwithin{equation}{section}

\usepackage[font=small,labelfont=bf,width=0.85\textwidth]{caption}

\expandafter\def\expandafter\bfseries\expandafter{\bfseries\ifmmode\else\boldmath\fi}
\expandafter\def\expandafter\mdseries\expandafter{\mdseries\ifmmode\else\unboldmath\fi}
\expandafter\def\expandafter\normalfont\expandafter{\normalfont\ifmmode\else\unboldmath\fi}

\RequirePackage{verbatim}

\makeatletter
\newwrite\bibinl@out
{\immediate\closeout\bibinl@out\@esphack}
\makeatother

\usepackage{fixmath}

\RequirePackage{amsmath}
\makeatletter
\newcommand{\brk@ord}{\bBigg@{0}}
\newcommand{\brk@ordl}{\mathopen\brk@ord}
\newcommand{\brk@ordr}{\mathclose\brk@ord}
\newcommand{\brk@ordm}{\mathrel\brk@ord}
\newcommand{\brk@var}{\brk@ord}
\newcommand{\brk@varl}{\left}
\newcommand{\brk@varr}{\right}
\newcommand{\brk@varm}{\mathrel\brk@var}
\newcommand{\brk@altname}[3]{\expandafter\def\csname#2\expandafter\@gobble\string#1\endcsname{#1[#3]}}
\newcommand{\brk@usearg}[3]{%
  \def\brk@star{*}\def\brk@blank{}\def\brk@arg{#1}%
  \ifx\brk@arg\brk@blank\def\brk@arg{brk@ord}\fi%
  \ifx\brk@arg\brk@star\def\brk@arg{brk@var}\fi%
  \csname\brk@arg #2\endcsname#3}

\newcommand{\DeclareMathBrackets}[3]{
  \newcommand{#1}[2][]{\brk@usearg{##1}{l}{#2}##2\brk@usearg{##1}{r}{#3}}
  \brk@altname{#1}{big}{big}\brk@altname{#1}{lr}{*}}
\newcommand{\DeclareMathBiBrackets}[4]{
  \newcommand{#1}[3][]{\brk@usearg{##1}{l}{#2}##2#3##3\brk@usearg{##1}{r}{#4}}
  \brk@altname{#1}{big}{big}\brk@altname{#1}{lr}{*}}
\newcommand{\DeclareMathBiMBracketsStar}[4]{
  \newcommand{#1}[3][]{\brk@usearg{##1}{l}{#2}##2\brk@usearg{##1}{m}{#3}##3\brk@usearg{##1}{r}{#4}}
  \brk@altname{#1}{bi}{big}}
\newcommand{\DeclareMathBiBracketsStar}[4]{
  \newcommand{#1}[3][]{\brk@usearg{##1}{l}{#2}##2\brk@usearg{##1}{}{#3}##3\brk@usearg{##1}{r}{#4}}
  \brk@altname{#1}{big}{big}}

\makeatother

\DeclareMathBrackets{\brk}{(}{)}
\DeclareMathBrackets{\sbrk}{[}{]}
\DeclareMathBrackets{\set}{\{}{\}}
\DeclareMathBrackets{\abs}{|}{|}
\DeclareMathBrackets{\eval}{.}{|}
\DeclareMathBrackets{\spn}{\langle}{\rangle}
\DeclareMathBiBrackets{\comm}{[}{,}{]}
\DeclareMathBiBrackets{\acomm}{\{}{,}{\}}
\DeclareMathBiBrackets{\gcomm}{[}{,}{\}}

\DeclareMathOperator{\tr}{tr}

\def\[{\begin{equation}}
\def\]{\end{equation}}

\providecommand{\href}[2]{#2}

\makeatletter
\def\mr@ignsp#1 {\ifx\:#1\@empty\else #1\expandafter\mr@ignsp\fi}%
\newcommand{\multiref}[1]{\begingroup%\let\protect\string%
\xdef\mr@no@sparg{\expandafter\mr@ignsp#1 \: }%
\def\mr@comma{}%
\@for\mr@refs:=\mr@no@sparg\do{\mr@comma\def\mr@comma{,}\ref{\mr@refs}}%
\endgroup}
\renewcommand{\eqref}[1]{(\multiref{#1})}
\makeatother

\makeatletter
\newcommand{\namedref}[2]{\hyperref[#2]{#1~\ref*{#2}}}
\newcommand{\secref}{\@ifstar{\namedref{Section}}{\namedref{Sec.}}}
\newcommand{\appref}{\@ifstar{\namedref{Appendix}}{\namedref{App.}}}
\newcommand{\tabref}{\@ifstar{\namedref{Table}}{\namedref{Tab.}}}
\newcommand{\figref}{\@ifstar{\namedref{Figure}}{\namedref{Fig.}}}
\makeatother

\usepackage{mathtools}
\DeclarePairedDelimiter\bra{\langle}{\rvert}
\DeclarePairedDelimiter\ket{\lvert}{\rangle}
\DeclarePairedDelimiterX\braket[2]{\langle}{\rangle}{#1 \delimsize\vert #2}

\providecommand{\hypersetup}[1]{}

\hypersetup{plainpages=false}
\hypersetup{pdfpagemode=UseNone}
\hypersetup{bookmarksnumbered=true}
\hypersetup{pdfstartview=FitH}
\hypersetup{colorlinks=false}
\hypersetup{citebordercolor={.5 1 .5}}
\hypersetup{urlbordercolor={.5 1 1}}
\hypersetup{linkbordercolor={1 .7 .7}}
\makeatletter
\let\@keywords\@empty
\let\@subject\@empty
\providecommand{\keywords}[1]{\gdef\@keywords{#1}}
\providecommand{\subject}[1]{\gdef\@subject{#1}}
\def\thetitle{\@title}
\def\theauthor{\@author}
\def\thesubject{\@subject}
\def\thedate{\@date}
\def\thekeywords{\@keywords}
\makeatother
\AtBeginDocument{
\hypersetup{pdftitle={\thetitle}}%
\hypersetup{pdfauthor={\theauthor}}%
\hypersetup{pdfsubject={\thesubject}}%
\hypersetup{pdfkeywords={\thekeywords}}%
}

\title{The Bethe ansatz for a new integrable open quantum system}
\author{ Marius de Leeuw, Chiara Paletta}

\begin{document}

\pdfbookmark[1]{Title Page}{title}
\thispagestyle{empty}

%\begingroup\raggedleft\footnotesize\ttfamily
%\arxivlink{yymm.nnnn}
%\par\endgroup

\vspace*{2cm}
\begin{center}%
\begingroup\Large\bfseries\thetitle\par\endgroup
\vspace{1cm}

\begingroup\scshape\theauthor\par\endgroup
\vspace{5mm}%

\begingroup\itshape
School of Mathematics
\& Hamilton Mathematics Institute\\
Trinity College Dublin\\
Dublin, Ireland
\par\endgroup
\vspace{5mm}

\begingroup\ttfamily
$\{$mdeleeuw,
palettac$\}$@maths.tcd.ie
\par\endgroup

\vfill

\textbf{Abstract}\vspace{5mm}

\begin{minipage}{12.7cm}
In this paper we apply the nested algebraic Bethe ansatz to compute the eigenvalues and the Bethe equations of the transfer matrix of the new integrable Lindbladian found in \cite{de2021constructing}. We show that it can be written as an integrable spin chain consisting of two interacting XXZ spin chains. We numerically compute the Liouville gap and its dependence on the parameters in the system such as scaling with the system length and interaction strength.

\end{minipage}

\vspace*{4cm}

\end{center}

\newpage

\tableofcontents

\newpage

%%%%%%%%%%%%%%%%%%%%%%%%%%%%%%%%%%%%%%%%%%%%%%%%%%%%%%%%%%%%%%%%%%%%%%%%%%%%%%%%
%%%%%%%%%%%%%%%%%%%%%%%%%%%%%%%%%%%%%%%%%%%%%%%%%%%%%%%%%%%%%%%%%%%%%%%%%%%%%%%%

\section{Introduction}

The coupling with the environment has a non-negligible influence on the dynamics of many-particle systems in both classical and quantum regimes.  Due to a vast number of applications such as, cold atoms, \cite{QNewtonCradle,gge-experiment1,ghd-experimental-atomchip,LL-GHD-exp}, electronic devices \cite{nava2021lindblad}, quantum circuits, \cite{lindblad-circuit,su2022integrable}  condensed matter, \cite{nava2021lindblad,Kavanagh:2021mra},  quantum optics \cite{alaeian2022exact,mitchison2020charging}, traffic models \cite{nava2022traffic}, the theory of open quantum system has become a topic of great interest in recent years. Within the Markovian approximation, an open system (reduced) density matrix evolves via the  Lindblad master equation, \cite{lindblad-eredeti,lindblad-intro,rivas2010markovian,manzano2018harnessing,breuer2002theory}

\begin{equation}
\dot{\rho}(t)=i[\rho,h]+\sum_a^K \left[
    \ell_a\rho \ell_a^\dagger-\frac{1}{2}\{\ell^\dagger_a \ell_a,\rho\}\right]=\mathcal{L}\rho(t) ,
    \label{lindblad}
\end{equation} 
where $h$ and $\ell_a$ are operators in a Hilbert space $\mathcal{H}$, $h$ is the Hamiltonian of the (closed) system  and $\ell_a$ is a set of jump operators describing the reservoirs responsible for the interaction with the environment. $\mathcal{L}$ is a super-operator acting on the space of bounded operators over the Hilbert space $\mathcal{H}.$\\ Due to the coupling with the environment, the dynamics of the system is non-trivial. The fixed points are non equilibrium steady states (NESS), \cite{wang2017theory}, the currents flowing through them continuously produce entropy.
\\
Given the difficulties to solve the Lindblad master equation analytically, particular attention was dedicated to \textit{exactly solvable} cases. Many different approaches have been suggested, such as: models solvable by free fermion techniques, \cite{third-quantization,third-quantization-2,katsura-lindblad-1,eric-lindblad}, boundary driven spin chains that allow the explicit computation of the NESS, \cite{ilievski2014exact,prosen-boundary-lindblad-1,prosen-boundary-lindblad-2,boundary-lindblad-mps,prosen-exterior-lindblad,enej-thesis,spin-helix-boundary,frassek2021exact}, triangular Lindblad superoperators for which the spectrum can be computed with exact techniques, \cite{marko-lindblad-1,japanok-lindblad}, models where the integrability is established separately for different subspaces of the full operator space, \cite{essler-piroli-lindblad}, strongly dissipative models for which the spectrum of the Lindbladian is derived perturbatively, \cite{PhysRevLett.126.190402}; and  Yang-Baxter integrable Lindblad systems where the reservoirs acts on the boundary of the spin chain \cite{prosen2013exterior} or with the reservoirs acting in the bulk, \cite{essler-prosen-lindblad,katsura-lindblad-2,essler-lindblad-review}.
\\
In this paper, we will focus on the last class, in particular one-dimensional quantum Yang-Baxter integrable models where the reservoir acts locally in the bulk in two adjacent sites of the spin chain. For those models, the superoperator $\mathcal{L}$ of $\eqref{lindblad}$ is one of the conserved charges $\mathbb{Q}s$ of an integrable model.
%\begin{align}
%\mathcal{L}=\mathbb{Q}_2.
%\end{align}
More precisely, the Lindblad superoperator is associated to a non-Hermitian Hamiltonian characterizing an integrable model.
The explicit expression of the Lindblad superoperator, for the case of one family of jump operator, is
\begin{equation}
  \mathcal{L}_{i,j}=-i{h_{i,j}}^{(1)}+i {h_{i,j}^{(2)}}^*+\left[
\ell_{i,j}^{(1)}  {\ell_{i,j}^{(2)}}^*-\frac{1}{2} \ell^{(1)\dagger}_{i,j} \ell^{(1)}_{i,j}
-\frac{1}{2}  {\ell^{(2)}_{i,j}}^T  {\ell^{(2)}_{i,j}}^*
\right]\,\,\,\in \mathcal{H}\otimes \mathcal{H}^*,\,\,\,j=i+1
\label{superop}
\end{equation}
The superscripts $^{(1)}$ and $^{(2)}$ identify in which space the operator acts.
\\
%For an integrable closed quantum system, the system evolves through the Generalized Gibbs ensamble, \cite{rigol-gge,JS-CGGE,rigol-quench-review} and the dynamics can be described via the Generalized Hydrodynamics, \cite{doyon-ghd,jacopo-ghd}. However some studies have been performed in weakly coupled regimes, \cite{doyon-weak-breaking,zala-tGGE,vasseur-breaking,integr-breaking-potential,space-time-inhom,alvise-js-adiabatic-formation,Lindblad-noise,jerome-atom-loss}, those procedures are not in general clear in the context of open quantum system.
%
A natural question to ask is whether some models exist for which the contribution of the environment preserves integrability and makes the system exactly solvable. This question was addressed in \cite{essler-prosen-lindblad}, where the authors presented an exactly solvable dissipative many-body quantum system. This model corresponds to a dephased spin-1/2 XX chain with a dissipative term and can be obtained by performing a unitary transformation on the Hubbard Hamiltonian with imaginary interaction strength. This allows to obtain the full spectrum via the Bethe Ansatz formalism. The question was also investigated in \cite{essler-lindblad-review,rowlands2018noisy,shibata2019dissipative} where the authors found the map (local or non-local) between some known integrable non-Hermitian Hamiltonian and the Lindblad formalism. These approaches try to map known integrable models to a Lindbladian. However, it is clearly desirable, to search and classify Yang-Baxter integrable Lindbladians more directly.  
\\
A new approach to deal exactly with this issue was put forward in \cite{de2021constructing}. The idea is to use the new boost authomorphism mechanism developed in \cite{Tetelman, HbbBoost,Loebbert:2016cdm,marius-classification-1,marius-classification-2,marius-classification-3,marius-classification-4}. By extending the boost approach to operators of Lindblad type we were able to initiate a systematic classification of Yang-Baxter integrable Lindblad systems.  The solution of this set give rise to \textit{new} Lindblad system corresponding to novel solutions of the Yang-Baxter equation. Furthermore, the method is complete and hence we found \textit{all} Yang-Baxter integrable Lindbladian corresponding to the types we considered.\\
Excitingly, we discovered a new model which has some very interesting features, which was called model B3. The Hamiltonian $h$ is a twisted spin-1/2 XX chain\footnote{the Hamiltonian can also be interpreted as the XX chain perturbed by a Dzyaloshinskii-Moriya interaction term, \cite{Dzyaloshinsky-Moriya-xxz}} and the jump operator $\ell$ is a nontrivial interaction, whose expression will be given in section \ref{modelexplicit}. Remarkably, the jump operator contains a coupling constant, which allows us to tune the strength of the environment. This model has a non-trivial, current carrying NESS and in some cases it also allows for a spin-helix state. This model seems to be the integrable Lindbladian which exhibits the richest physical properties. \\
For this reason, it is important to be able to study this model in more detail. In this paper we take a first step towards this by performing the algebraic Bethe Ansatz which allows us to find all the eigenvectors and eigenvalues of the superoperator. 
%
%In section \ref{4dint}, we give a 4-D interpretation as two spin-1/2 XXZ chains interacting. The anisotropy $\Delta$ is proportional to the coupling constant of the model.\\
%
Since our local Hilbert space is four dimensional, we need to apply the so-called nested algebraic Bethe ansatz technique, see \cite{Levkovich-Maslyuk:2016kfv,Slavnov:2019hdn} for recent reviews. We found that this approach for our model closely mirrors of the ones for the Hubbard model, \cite{ramos1997algebraic,martins1998quantum} and the $\text{AdS}_5 \times \text{S}_5$  for bound states, \cite{arutyunov2009bound}.
\\
%\comC{I don't know if we should cite some other papers here, for example:\\
%two $\eta$ deformations of $\text{AdS}_5 \times \text{S}_5$ \cite{seibold2021bethe} or \cite{beisert20082}. Maybe also Sfondrini and Frolov works?
%}\\
%We identified $B_1, B_2$ and $B_3$ as the creation operators of the theory. The first two create electrons with spin up and down, while the third one creates a pair. We explicitly constructed the eigenvectors of the model for one $\ket{1}=F^a B_a(v_1)\ket{0}$ and two magnons (Appendix \ref{2partstates}) $\ket{2}=F^{ab} B_{a}(v_1)B_{b}(v_2)\ket{0}$. The nesting is manifest, in fact $F$ should be the eigenvector of the transfer matrix of another integrable model (twisted 6-vertex type). 
%
We give an interpretation for model B3 as two coupled spin-1/2 XXZ chains. The anisotropy $\Delta$ is proportional to the coupling constant of the model. In view of this, we identified three types of creation operators $B_1, B_2$ and $B_3$. The first two create electrons with spin up and down, while the third one creates an electron pair. 
We find that a state is characterized by the length $L$ of the spin chain as well as
\begin{align}
\nonumber
&M=\#\, \text{electrons }\\ % \{v\}=(v_1,\dots,v_M)\\
\nonumber
&N=\#\, \text{number of spins up }%\{w\}=(w_1,\dots, w_N)
\end{align}
The main result of the paper is the expression of the eigenvalue of the transfer matrix for a chain of length $L$,\\
%\comC{to be changed}\textcolor{blue}{done, shifted}
\begin{align}
\nonumber\frac{\lambda(u)}{\mathcal{N}}=&\prod _{i=1}^M \frac{\sinh \left(u-\mathtt v_i+\frac{3 \Psi }{2}\right)}{\sinh\left(u-\mathtt v_i+\frac{\Psi }{2}\right)}+\\
\nonumber&\frac{\lambda_{6V}(u)}{\left(-e^{2 i \phi }\right)^{L-M-N}} \prod _{i=1}^M  \frac{\sinh \left(u-\mathtt v_i-\frac{\Psi }{2}\right)}{\sinh\left(u-\mathtt v_i+\frac{\Psi }{2}\right)} \prod _{j=1}^L \frac{  \sinh \left(u-\mathtt u_j+\psi \right)}{i\,e^{\Psi +i \phi }\sinh\left(u-\mathtt u_j\right)}+\\
&\prod _{i=1}^M \frac{\cosh \left(u-\mathtt v_i-\frac{3 \Psi }{2}\right)}{\cosh \left(u-\mathtt v_i-\frac{\Psi }{2}\right)}\prod _{i=1}^L \frac{ \coth \left(u-\mathtt u_i\right) \sinh \left(u-\mathtt u_i+\Psi \right)}{e^{2 \Psi }\,\cosh \left(u-\mathtt u_i-\Psi \right)},
\end{align}

$\mathcal{N}=\left(-e^{2 i \phi }\right)^{M-N} \left(-e^{\Psi -i \phi }\right)^M$ and for the nested chain
\begin{align}
\frac{\lambda _{6V}(u)}{\left(-e^{2 i \phi }\right)^{L-M}}=&\prod _{i=1}^M-\frac{ \sinh \left(2 \left(u-w_i+\Psi \right)\right)}{e^{2 i \phi }\,\sinh\left(2 \left(u-w_i\right)\right)}+\\
\nonumber&\left(-e^{2 i \phi }\right)^{L-M} \prod _{i=1}^N -\frac{ \sinh \left(2 \left(u-w_i-\Psi \right)\right)}{e^{2 i \phi } \sinh\left(2 \left(u-w_i\right)\right)}\prod _{i=1}^M -\frac{e^{2 i \phi } \sinh \left(2 u-2 \mathtt v_i+\Psi \right)}{\sinh\left(2 u-2 \mathtt v_i-\Psi \right)}.
\end{align}
The rapidities $\{v\}$ and $\{w\}$ of the particles in the model obey the sets of Bethe equations, for $j=1,\dots,M$
\begin{align}
&\prod _{i=1,i\neq j}^M \frac{\sinh\left(\mathtt v_i-\mathtt v_j-\Psi \right)}{\sinh\left(\mathtt v_i-\mathtt v_j+\Psi \right)}=\prod _{i=1}^L \frac{1}{i\,e^{\Psi +i \phi }}\frac{  \sinh\left(\mathtt u_i-\mathtt v_j-\frac{\Psi }{2}\right)}{\sinh \left(\mathtt u_i-\mathtt v_j+\frac{\Psi }{2}\right)} \prod _{i=1}^N \frac{\sinh \left(2 \mathtt v_j-2 w_i+\Psi \right)}{\sinh\left(2 \mathtt v_j-2 w_i-\Psi \right)}
\end{align}
and the auxiliary Bethe equations, for $j=1,\dots,N$
\begin{equation}
\prod _{i=1,i\neq j}^N \frac{\sinh \left(2 \left(w_i-w_j-\Psi \right)\right)}{\sinh\left(2 \left(w_i-w_j+\Psi \right)\right)}=\left(-e^{2 i \phi }\right)^{L}\prod _{i=1}^M \frac{\sinh\left(2 \mathtt v_i-2 w_j-\Psi \right)}{\sinh\left(2 \mathtt v_i-2 w_j+\Psi \right)},
\end{equation}
We have checked explicitly that the Bethe Ansatz gives the correct result for spin chains of small length and up to three particles state.\\
%\comC{add Bethe equations}\\
Finally, in section \ref{numerics}, we numerically study the eigenvalues for the transfer matrix of a chain of $L=8$ in different regimes of the coupling constants. The second biggest real part of the eigenvalues of the superoperator is called spectral gap. In finite systems, it is equivalent to the inverse of the relaxation time, \cite{vznidarivc2015relaxation,essler-prosen-lindblad}. We analysed the scaling of the spectral gap with the dimension of the chain, called dynamical exponent and we found that it scales as $1/L^2$. We plotted the dependence of the gap on the coupling constant and we found that $\text{gap}\sim \tanh \psi$, which is the coupling constant of the model. 

\section{Model B3}
\label{modelexplicit}

In this section, we give the explicit expression of the Hamiltonian $h$, the jump operator $\ell$, the superoperator $\mathcal{L}$, \eqref{superop} and the $R$-matrix characterizing our integrable model. We will also discuss its interpretation as a regular integrable model with a local Hilbert space of dimension 4.

\subsection{Hamiltonian and jump operator}
Model B3 is characterized by the following constant Hamiltonian $h$ \begin{align}
&\frac{h_{j,j+1}}{\alpha}=
 \left(
\begin{array}{cccc}
 0 & 0 & 0 & 0 \\
 0 & 0 & e^{i \phi } & 0 \\
 0 & e^{-i \phi } & 0 & 0 \\
 0 & 0 & 0 & 0 \\
\end{array}
\right) = e^{i \phi }\,  \sigma^+_j \sigma^-_{j+1}  + e^{-i \phi }\,  \sigma^-_j \sigma^+_{j+1} ,
\label{hb3}
\end{align}
and jump operator $\ell$ 
\begin{align}
\frac{\ell_{j,j+1}}{\beta}=
&\left(
\begin{array}{cccc}
 i \sinh \psi  & 0 & 0 & 0 \\
 0 & i \cosh \psi & e^{-\psi +i \phi } & 0 \\
 0 & e^{\psi -i \phi } & -i \cosh \psi  & 0 \\
 0 & 0 & 0 & i \sinh \psi  \nonumber\\
\end{array}
\right)  \\
& =i \sinh \psi \,(1+2 n_j n_{j+1})+i \,(e^{-\psi } n_j - e^{\psi } n_{j+1})+ e^{-\psi +i \phi } \sigma_j^+\sigma_{j+1}^-+e^{\psi -i \phi } \sigma_j^-\sigma_{j+1}^+,
\label{lb3}
\end{align}
where $\alpha=\cosh ^2\psi \, \text{sech}\,2 \psi $ and $\beta=-i \sqrt{\tanh \psi \, \text{sech}\,2 \psi }$. Both the Hamiltonian and the jump operator have range two and describe nearest neighbour interactions. Notice that we have parameterized $\gamma = \tanh \,\psi$ compared to \cite{de2021constructing}.

\subsection{4D spin chain interpretation}
\label{4dint}

In order to construct the superoperator, we need to work in the Fock-Liouville space $\mathcal{H}\otimes \mathcal{H}^*$, where $\mathcal{H}\equiv \mathbb{C}^2 \otimes \mathbb{C}^2$. In what follows, a general operator $A_j$ acts on $\mathcal{H}$, while $\tilde{A}$ acts on $\mathcal{H}^*$. Seen in this way our superoperator defines a spin chain Hamiltonian $\mathbb{H}$ on a spin chain with local dimension 4. 
\\
For this reason, we can decompose it in terms of two sets of standard fermionic operators $\sigma,\tilde{\sigma}$.
Explicitly, the action of the set of fermionic operators is
\begin{align}
&\sigma_j^\pm=\sigma^\pm \otimes 1 \otimes 1 \otimes 1 ,\\
&\sigma_{j+1}^\pm=1 \otimes  1 \otimes \sigma^\pm \otimes  1,
\\
&\tilde\sigma_{j}^\pm=1 \otimes \sigma^\pm \otimes  1 \otimes  1,
\\
&\tilde\sigma_{j+1}^\pm=1 \otimes  1 \otimes  1 \otimes \sigma^\pm,
\end{align}
where $\sigma^+=\frac{\sigma_x+ i \sigma_y}{2}$ and $\sigma^-=\frac{\sigma_x- i \sigma_y}{2}$.\\
The Lindblad superoperator \eqref{superop} is then
\begin{equation}
 \mathbb{H}_{ij} \equiv \mathcal{L}_{i,j}=-i{h_{i,j}}+i {\tilde h_{i,j}}+
\ell_{i,j}  \tilde \ell_{i,j}-\frac{1}{2} \ell^{\dagger}_{i,j} \ell_{i,j}-\frac{1}{2}  {\tilde \ell_{i,j}}  {\tilde \ell_{i,j}}^*,
\end{equation}
and can be rewritten as
\begin{align}
\mathbb{H}_{i,j}={J_{i,j}}+ \tilde J_{i,j}+ { \ell_{i,j}}  {\tilde \ell_{i,j}},
  \label{sup}
\end{align}
where
\begin{align}
J_{i,j}=i \,h_{i,j}-\frac{1}{2} \ell^{\dagger}_{i,j} \ell_{i,j}.
\end{align}
In terms of oscillators, for the model B3 characterized we obtain
\begin{align}
&J_{i,j}=i e^{i \phi } (\tanh \psi -1)\sigma_i^+ \sigma_j^--ie^{-i \phi } (\tanh \psi +1)\sigma_i^- \sigma_j^+  - \tanh \psi \,(n_i-n_j)^2,\label{Jij}\\
&\tilde J_{i,j}= -i e^{-i\phi} \, (\tanh \psi -1)\tilde\sigma_i^+ \tilde\sigma_j^-+i e^{i \phi }(\tanh \psi +1) \tilde\sigma_i^- \tilde\sigma_j^+-\tanh \psi ( \tilde n_i -\tilde n_j)^2,
\end{align}
we notice that the coefficients of the terms in $J$ are the complex conjugate of the ones in $\tilde J$. The term $\ell_{ij}\tilde \ell_{ij}$ in \eqref{sup} is
\begin{align}
\ell_{ij}\tilde \ell_{ij}=
\frac{\tanh\psi}{\cosh2\psi}(2 \sinh \psi \, n_i n_j+e^{-\psi }  n_i-i e^{-\psi +i \phi }\sigma_i^+ \sigma_j^--i e^{\psi -i \phi }\sigma_i^- \sigma_j^+-e^{\psi }\, n_j)(c.c)
\label{ells}
\end{align}
where $(c.c)$ has the tilded operators and the coefficients are complex conjugate. We notice that when we send $\psi\rightarrow0$ our Hamiltonian $\mathbb{H}$ simply decomposes into two independent XX spin chains.
\\
Interestingly, the expression \eqref{Jij} of $J_{i,j}$  corresponds, up to a twist and a renormalization, to the Hamiltonian of the XXZ chain
\begin{align}
U_i J_{i,j} U_i^{-1} \,=\,(h_{XXZ})_{i,j}\,=\, \sigma_i^x \sigma_j^x+\sigma_i^y \sigma_j^y+\Delta \sigma_i^z \sigma_j^z,
\end{align}
where $\Delta=\pm i \sinh \psi $. The twist $U_i$ is
\begin{align}
U_i=\left(
\begin{array}{cc}
 1 & 0 \\
 0 & \pm e^{-\psi +i \phi } \\
\end{array}
\right).
\end{align}
To summarize, we see that interpreted as a normal spin chain, our model B3 corresponds to two coupled XXZ chains with interaction between the two given by the $\ell$s term \eqref{ells}. Applying the same twist $U$ on the jump operator $\ell$ we get
\begin{align}
U_i \ell_{i,j} U_i^{-1}=\mathcal{M} \left(
\begin{array}{cccc}
 \tanh \psi  & 0 & 0 & 0 \\
 0 & 1 & \mp {i\, \text{sech}\psi} & 0 \\
 0 & \mp i  \, \text{sech}\psi  & -1 & 0 \\
 0 & 0 & 0 & \tanh \psi  \\
\end{array}
\right),
\end{align}
where $\mathcal{M}=(\tanh \psi +\coth \psi )^{-1}$. Hence we see that, in contrast to the Hubbard model, the coupling constant between the two independent chains is related to the interaction strength $\Delta$ in the individual spin chains.

\subsection{$R$-matrix}
\label{rmatrix}

The characterizing object in an integrable model is the $R$-matrix, $R(u)\in {\rm End}(\mathcal{H}\otimes \mathcal{H})$ which is a solution of the Yang-Baxter equation
\begin{equation}
R_{12}(v-u)R_{13}(v)R_{23}(u)=R_{23}(u)R_{13}(v)R_{12}(v-u),
\end{equation}
on $\mathcal{H}\otimes \mathcal{H}\otimes \mathcal{H}$, the subscripts denote which of the spaces $R$ acts on.
\\
The entries of the $R$-matrix characterizing model B3 are
\begin{align}
\nonumber &R_1^1= R^6_6= R^{11}_{11}= R^{16}_{16}= 1,\\
\nonumber &\frac{R^2_2}{e^{i \phi }}= -e^{i \phi }R^3_3 = -e^{i \phi }R^8_8 = \frac{R^{12}_{12}}{e^{i \phi }}= \frac{i (\tanh \psi +1)}{\coth (u-\psi )+\tanh \psi },\\
\nonumber & -e^{i \phi } R^5_5= \frac{R^9_9}{e^{i \phi }}= \frac{R^{14}_{14}}{e^{i \phi }}=-e^{i \phi } R^{15}_{15} = \frac{i (\tanh \psi -1)}{\coth (u-\psi )+\tanh \psi },\\
\nonumber &R^2_5= R^5_2= R^{15}_{12}= R^{12}_{15}= R^8_{14}= R^{14}_8= R^9_3= R^3_9= \text{sech} u \cosh \psi,\\
\nonumber & e^{i \phi } R^4_7= -\frac{R^4_{10}}{e^{i \phi }}= \frac{R^{13}_{10}}{e^{i \phi }}= -e^{i \phi }R^{13}_7 = -i \text{sech }u \cosh \psi  \sinh (u-\psi ) \text{sech}(u+\psi ),\\
\nonumber &-e^{i \phi }R^7_4 = \frac{R^{10}_4}{ e^{i \phi }}= i e^{2 \psi } \text{sech }u \cosh \psi  \sinh (u-\psi ) \text{sech}(u+\psi ),\\
\nonumber &e^{i \phi }R^7_{13} = -\frac{R^{10}_{13}}{e^{i \phi }}= i e^{-2 \psi } \text{sech }u \cosh \,\psi  \sinh (u-\psi ) \text{sech}(u+\psi ),
\\
\nonumber &R^{10}_7= R^7_{10}= \frac{1}{2} \text{sech }u (\cosh \,\psi +\cosh (3 \psi )) \text{sech}(u+\psi ),\\
\nonumber &e^{2 i \phi }R^7_7 = \frac{R^{10}_{10}}{e^{2 i \phi }}= \frac{R^{4}_{4}}{e^{2 i \phi }}= \frac{R^{13}_{13}}{e^{-2 i \phi }}=\tanh \,u \sinh (u-\psi ) \text{sech}(u+\psi ),\\
\nonumber &R^4_{13}= -e^{u-\psi } \text{sech }u \cosh \psi  (\sinh (u-\psi ) \text{sech}(u+\psi )-1),\\
\nonumber &R^{13}_4= e^{\psi -u} \text{sech }u \cosh \psi  (\sinh (u-\psi ) \text{sech}(u+\psi )+1).
\end{align}
where $R_i^j$ is the element in the $i$th-row and $j$th-column and for simplicity we omitted the dependence on the shifted spectral parameter. Explicitly
\begin{align}
R_i^j=R_i^j(u-\psi).
\end{align}
This matrix is the same as the one given in the Supplemental material of the letter\footnote{We would like to thank the authors of \cite{su2022integrable} for pointing out the typos in the $R$-matrix.} \cite{de2021constructing} by considering the rescaling of the spectral parameter and the parametrization of the coupling constant 
\begin{align}
& u \to \frac{2\,u}{ \left(\gamma ^2+1\right)},
&&\gamma = \tanh \,\psi.
\end{align}
Notice that this reparametrization maps strong coupling to strong coupling. Hence strong coupling is $\psi\rightarrow \infty$ and weak coupling corresponds to $\psi \rightarrow 0$.

\section{Diagonalization of the transfer matrix}
\label{nestingBA}
This section contains the main result of the paper. The aim of the Algebraic Bethe ansatz  is to find the eigenvalues of the transfer matrix \cite{faddeev1996algebraic}. From this, in a systematic way, one can construct the eigenvalues of the tower of the conserved charges characterizing the integrable model. We define the monodromy, the transfer matrix and the reference state.  By using the RTT relation, we give the commutation relations between the entries of the monodromy matrix and their interpretation. Then we explicitly compute the eigenvalue of the transfer matrix and the Bethe equations, used to determine the momenta of the particles involved in the theory, for a state of one and two magnons and explain how to generalize the result to an arbitrary number of particles.

\subsection{Monodromy and transfer matrix. Definitions}
To define the monodromy matrix $T_a(u)$ for a spin chain of length $L$, we need to introduce an auxiliary Hilbert space $V_a$
\begin{equation}
T_a(u)=\prod_{i=1}^L R_{a i}(u_i-u-b),\,\,\,\,\,T_a(u)\in  V_a \otimes\underbrace{V \otimes \dots \otimes V}_{L-times},
\label{monodromy1}
\end{equation}
$u_i$ is the set of inhomogeneities of the chain and $b$ is a constant.
The transfer matrix is defined as the partial trace ($\tr_a$) over the auxiliary space of the monodromy matrix,
\begin{equation}
T(u)=\tr_a T_a (u),\,\,\,\,\,T(u)\in  \underbrace{V \otimes \dots \otimes V}_{L-times}.
\label{transfer1}
\end{equation} 
This matrix generates all the conserved charges characterizing the integrable models, in particular the charge $\mathbb{Q}_2\equiv\mathcal{L}$ which will be identified as the logarithmic derivative of the transfer matrix.
\\For a regular, homogeneous model\footnote{For these models $R(0)=P$, where $P$ is the permutation operator acting on two copies of $V$.}, the $R$-matrix is related to a range-2 charge
\begin{equation}
\frac{d}{d u} R(u)|_{u\to 0}=P\, \mathbb{Q}_2,
\end{equation}
which corresponds to the Hamiltonian.
\\
\subsection{Monodromy and transfer matrix. Constructions}
The monodromy matrix \eqref{monodromy1} in the auxiliary space takes the form of a $4\times 4$ matrix 
\begin{equation}
T_a(u)=\left(
\begin{array}{cccc}
 T_{00} & B_1 & B_2 & B_3 \\
 C_1 & T_{11} & T_{12} & T_{13} \\
 C_2 & T_{21} & T_{22} & T_{23} \\
 C_3 & T_{31} & T_{32} & T_{33} \\
\end{array}
\right),
\label{monodromy}
\end{equation}
where the entries of this matrix are operators acting on the physical space $\underbrace{V \otimes \dots \otimes V}_{L-times}$. For simplicity, we omitted the $(u)$ dependence from all the entries.\\
The transfer matrix \eqref{transfer1} is then
\begin{equation}
T(u)=\sum_{i=0}^3 T_{ii}(u).
\end{equation}
\\
The monodromy matrix and the $R$-matrix satisfy the fundamental commutation relations, also known as the RTT-relations,
\begin{equation}
R_{ab}(v-u)T_a(u)T_b(v)=T_b(v)T_a(u)R_{ab}(v-u).
\label{rtt}
\end{equation}
The space where this matrix acts is  $V_a \otimes V_b \otimes \underbrace{V\otimes \dots \otimes V}_{L \,\text{times}}$, with $V_a$ and $V_b$ auxiliary spaces.
\\
By plugging the expression of the $R$-matrix and the monodromy matrices given respectively  in section \ref{rmatrix} and \eqref{monodromy}, it follows that
\begin{equation}
[B_i(u),B_i(v)]=0,\,\,\,\,\,i=1,2,3,
\label{commrels}
\end{equation}
which gives the immediate interpretation: $B_1$ and $B_2$ (and also $B_3$) are the creation operators for our theory.

\subsection{The reference states and the action of the transfer matrix}
\label{therefstate}
Since model B3 preserves the spin, a good choice for the reference state is
\begin{equation}
\ket{0}=\bigotimes_{i=1}^L\left(
\begin{array}{c}
 1 \\
 0 \\ 0 \\ 0 \\
\end{array}
\right).
\end{equation}
%\comC{I really like this command underbrace, am I over using it? ahah}\\
The action of the elements of the transfer matrix  on the reference state $\ket{0}$, by fixing the constant $b=\psi$ is
\begin{equation}
T_{00}(u)\ket{0}=\ket{0},
\label{T00vacuum}
\end{equation}
\begin{equation}
T_{11}(u)\ket{0}=\prod_{i=1}^L \frac{ \sinh \left(u-u_i+\psi \right)}{i\,e^{\psi +i \phi }\,\cosh \left(u-u_i\right)}\ket{0},
\label{T11vacuum}
\end{equation}
\begin{equation}
T_{22}(u) \ket{0}=\prod_{i=1}^L   \frac{i\,\sinh \left(u-u_i+\psi \right)}{e^{\psi - i \phi }\,\cosh \left(u-u_i\right)}\ket{0},
\label{T22vacuum}
\end{equation}
\begin{equation}
T_{33}(u) \ket{0}=\prod_{i=1}^L e^{-2 \psi }\frac{\sinh \left(u-u_i+\psi \right)}{\cosh \left(u-u_i-\psi \right)} \tanh \left(u-u_i\right)  \ket{0},
\label{T33vacuum}
\end{equation}
and the following annihilation identities hold
\begin{equation}
C_1 \ket{0}=C_2 \ket{0}=C_3 \ket{0}=0,\,\,T_{ab}\ket{0}=0\,\, (a\neq b =1,2).
\end{equation}
$\{u_i\}$, already introduced in \eqref{monodromy1}, are the set of inhomogeneities of the spin chain. From now on, we will refer to it as \textit{main} spin chain, for a reason that will be clear in the following.\\
The action of the transfer matrix on the vacuum is
\begin{align}
 T(u) \ket{0}=\prod_{i=1}^L &\bigg[e^{-2 \psi } \sinh \left(u-u_i+\psi \right) \bigg(\frac{\tanh \left(u-u_i\right)}{\cosh \left(u-u_i-\psi \right)}-\frac{2 e^{\psi } \sin \phi }{\cosh \left(u-u_i\right)}\bigg)+1\bigg]\ket{0}.
\end{align}
Due to the commutation relations \eqref{commrels}, an excited state can be constructed by acting with the operators $B_1$, $B_2$ and $B_3$ on the vacuum. As an example, a state of two particle with rapidities $v_1$ and $v_2$ is
\begin{equation}
B_1(v_1)B_2(v_2) \ket{0}.
\end{equation}
In what follows, we will explicitly construct states of one and two magnons that are also eigenstates of the transfer matrix.
\\
To understand if a state is an eigenstate of the transfer matrix, we need to find the commutation relations between $T_{ii}(u)$ and the $B$s operators and then act with them on the vacuum via \eqref{T00vacuum}-\eqref{T33vacuum}. The commutation relations can be found from the RTT \eqref{rtt} and we will explicitly give them in what follows. Furthermore, the condition that a state is an eigenstate will fix a constraint on the rapidities $v_i$ of the particles, the Bethe equations.

\subsection{Commutation relations: here comes the nesting}
%\comC{referring to "Here comes the sun" from the Beatles, I like it, but if you don't like feel free to remove}\\
Before giving the commutation relations between $T_{ii}(u)$ and the $B$s, we want to focus on the meaning of the operator $B_3$.
\subsubsection{Commutation relation between $B$s}
From the RTT-relations \eqref{rtt}, it follows that
\begin{equation}
B_{\alpha }(u) B_{\beta }(v)=B_{\delta}(v) B_{\gamma }(u) r_{\alpha \,\beta  }^{\gamma \,\delta}(v-u)-\epsilon _{\alpha ,\beta } \eta(u-v)\left(B_3(v) T_{00}(u)-B_3(u) T_{00}(v)\right),
\label{BBcomm}
\end{equation}
where $\alpha, \beta = 1, 2$ and
\begin{equation}
\eta(u)=i e^{2 \psi } \cosh \psi \, \text{csch}(u-\psi ),\,\,\,\epsilon=(\epsilon_{11},\epsilon_{12},\epsilon_{21},\epsilon_{22})=(0,e^{-i \phi },-e^{i \phi },0).
\label{eta}
\end{equation}
The elements $r_{\alpha \,\beta  }^{\gamma \,\delta}(u)$ can be written in matrix form, 
\begin{align}
&r(u)=r_{\alpha \,\beta  }^{\gamma \,\delta}(u) e_\gamma^\alpha \otimes e_\delta^\beta,
&&r(u)=\left(
\begin{array}{cccc}
 1 & 0 & 0 & 0 \\
 0 & b(u)e^{-2 i \phi} & a(u) & 0 \\
 0 &  a(u) &b(u)e^{2 i \phi} & 0 \\
 0 & 0 & 0 & 1 \\
\end{array}
\right),
\label{rnested}
\end{align}
where $a(u)=\frac{\sinh (2 \psi )}{\sinh (2 (u+\psi ))}$ and $b(u)=\frac{\sinh (2 u)}{\sinh (2 (u+\psi ))}$.\\
It is easy to show that $r(u)$ is an $R$-matrix on a spin-1/2 chain of twisted 6-vertex type.
\\
This is the first insight of why the Bethe ansatz is called \textit{nested}: in the commutation relations involving different type of particles the $r$-matrix of a lower dimensional spin chain appears. The same $r$ also appears from the RTT relations involving $T$ and $B$ operators.\\
Note that \eqref{commrels} can also be written as \eqref{BBcomm} with $\alpha=\beta$.
From \eqref{BBcomm}, we can give an interpretation for $B_1$, $B_2$ and $B_3$. The commutation relations between two fields of type $B_1$ and $B_2$ generate the operator  $B_3$. One can consider the operators $B_1$ and $B_2$ as creation of a particle of spin up and down respectively, while $B_3$ is responsible for the creation of a pair.\\
The $r$ matrix \eqref{rnested} is a twisted version of the ones that appear in the Hubbard model and in $AdS_5 \times S^5$, \cite{ramos1997algebraic,martins1998quantum, arutyunov2009bound}.
\subsubsection{Commutation relations between $T_{ii}$ and $B$s}
As mentioned, we need to solve the eigenvalue problem
\begin{equation}
T(u)\ket {M\{v\}}=
\sum_{i=0}^3 T_{ii} (u)\ket {M\{v\}}=\lambda(u)\ket {M\{v\}},
\end{equation}
where $\ket{M\{v\}}$ is a generic state of $M$ excitations with rapidities $\{v\}$. First, we need to find the commutation relations between $T_{ii}$s and $B$s.\\
From the RTT one gets 256 relations, but not all of them are already in a usable form. In particular, we want the right hand side to be normal ordered and have annihilation and diagonal operators on the right most side.\\
However, taking linear combinations gives us the wanted structure
\begin{equation}
T_{ii}(u)B_\alpha(v)=a_1 B_\alpha(v)T_{ii}(u)+a_2 B_\alpha(u) T_{ii}(v)+\dots,
\label{comrelcond}
\end{equation}
the dots ($"\dots"$) contains terms that either annihilate the reference state (for example in the right there is $C_i$) or acts diagonally on it.\\
Instead of trying the most general linear combination, we first tried to impose that the structure of our commutator relations is the same as the ones in \cite{ramos1997algebraic,martins1998quantum, arutyunov2009bound}. This drastically simplifies the problem and we find
\begin{align}
&T_{00}(u)B_\alpha(v)=&&\theta_\alpha B_\alpha(v)T_{00}(u)+\rho_\alpha B_\alpha(u)T_{00}(v)\label{T00B}\\
\nonumber&T_{\alpha \alpha^\prime}(u)B_\beta (v) =&&  \alpha_{\alpha}B_{\gamma}(v)T_{\alpha \tau}(u)r_{\alpha^\prime \beta}^{\tau \gamma}(v-u)-\psi_\alpha B_{\alpha^\prime} (u) T_{\alpha \beta}(v)-\\
& && (\upsilon T_{\alpha 3} (u)T_{00}(v)+\beta_\alpha B_3 (u)C_\alpha(v)+\gamma_\alpha B_3 (v) C_\alpha(u))\epsilon_{\alpha^\prime \beta}\label{Taa}\\
&T_{33}(u)B_\alpha(v) =&& \zeta_{1,\alpha} B_\alpha(v)T_{33}(u)+\zeta_{2} B_3(v)T_{3\alpha}(u)+\zeta_3 B_3(u)T_{3 \alpha}(v)+\eta \epsilon_{\gamma \eta}T_{\gamma 3}(u)T_{\eta \alpha}(v)\label{T33B},
\end{align}
for simplicity we omitted the spectral dependence of the coefficients, which is $\omega=\omega(u-v)$ for all of them.\\
Remarkably, in \eqref{Taa} we again notice the $r$-matrix of twisted 6-vertex type given in \eqref{rnested}. This is another strong insight of why the Bethe ansatz is called nested and the role played by this matrix will be clear in the next paragraph. In fact, in order to solve the Bethe ansatz for  model B3, we first need to solve the Bethe ansatz for the integrable model characterized by the $r$-matrix of twisted 6-vertex type. We also mention that the commutation relations found here are independent on the choice of the constant $b$ in \eqref{monodromy1}.
\\The coefficients in the commutation relations are
\begin{align}
 & \theta_1=-\frac{i e^{\psi +i \phi } \cosh (u+\psi )}{\sinh u},
&& \rho_1=\frac{i \cosh \psi  e^{\psi +i \phi }}{\sinh u},
&& \theta_1=-e^{2 i \phi } \theta_2,
&& \rho_1=-e^{2 i \phi } \rho_2,
\label{theta}
\end{align}
\begin{align}
\nonumber  & \alpha_1=i  e^{\psi +i \phi }  \cosh \psi \,(\coth u-\tanh \psi ),
&& \psi_1=i e^{\psi +i \phi }  \cosh \psi  \, \text{csch}u,\\
\nonumber & \upsilon = i e^{2 \psi } \cosh \psi  \, \text{csch}(u-\psi ),
&&\beta_1= -e^{3 \psi +i \phi }\cosh ^2\psi\,  \text{csch}u \, \text{csch}(u-\psi ),\\
&\gamma_1= e^{3 \psi +i \phi } \cosh \psi \, \text{csch}u \, \coth (u-\psi )
\label{alpha}
\end{align}
\begin{align}
&\alpha_1=-e^{2 i \phi } \alpha_2,
&&\psi_1=-e^{2 i \phi } \psi_2,
&&\beta_1=-e^{2 i \phi } \beta_2,
&&\gamma_1=-e^{2 i \phi } \gamma_2
\label{condalpha}
\end{align}
\begin{align}
\nonumber &\zeta_{1,1}=\frac{i e^{\psi +i \phi } \cosh (u-2 \psi )}{\sinh (u-\psi )},
&&\zeta_2=\frac{e^{2 \psi } \cosh \psi  \cosh (u-2 \psi )}{\sinh u \sinh (u-\psi )},\\
&\zeta_3=-  e^{u+2 \psi }\frac{\cosh \psi}{\sinh (u-\psi )} \left(\frac{\cosh (u-2 \psi )}{\sinh u}-1\right),
&&\zeta_{1,1}=-e^{2 i \phi } \zeta_{1,2},
\label{zeta}
\end{align}
where the dependence on the spectral parameter is $\omega=\omega(u)$. $\eta$ and $\epsilon$ in  \eqref{T33B} are the same as the ones in \eqref{eta}.

\subsection{One particle state}
\label{onepstate}
This section and the appendix \ref{2partstates} will help to understand the general derivation for arbitrary number of particles.\\
One magnon can be created either by $B_1$ or $B_2$, so the one particle state is a linear combination of these two with weight $F^a$
\begin{equation}
\ket{1\{v\}}=F^a B_a(v) \ket 0,
\end{equation}
where we sum over the repeated index ($a=1,2$), and $\{v\}=v$ is the rapidity of the magnon.\\
By using \eqref{T00B}-\eqref{T33B} and \eqref{T00vacuum}-\eqref{T33vacuum}, the action of the transfer matrix on one-particle state is
\begin{equation}
T_{00}(u) \ket {1 \{v\}}= \theta_a(u-v) F^a B_a(v) T_{00}(u) \ket{0}+\dots=\theta_a(u-v)  F^a B_a(v) \ket 0+\dots,
\label{T00one}
\end{equation}
and similarly for $T_{33}(u)$
\begin{equation}
T_{33}(u) \ket {1 \{v\}}=\zeta_{1,a}(u-v)\prod_{i=1}^L e^{-2 \psi }\frac{\sinh \left(u-u_i+\psi \right)}{\cosh \left(u-u_i-\psi \right)} \tanh \left(u-u_i\right) F^a\,B_a(v) \ket 0+\dots.
\label{T33one}
\end{equation}
The terms $T_{11}$ and $T_{22}$ require particular analysis. First, it is convenient to write the relations \eqref{T11vacuum} and \eqref{T22vacuum} in the form
\begin{align}
T_{\alpha \tau}(u)\ket{0}=\delta_{\alpha \tau}\prod_{i=1}^L \,(-e^{2 i \phi})^{\alpha-1}f(u,\{u_i\})\ket{0},
\label{Taaone}
\end{align}
where $f(u,\{u_i\})= \frac{ \sinh \left(u-u_i+\psi \right)}{i\,e^{\psi +i \phi }\cosh \left(u-u_i\right)}$.\\
The action of $T_{11}+T_{22}$ is
\begin{align}
\nonumber  T_{\alpha \alpha}(u)\ket{1}=& F^a T_{\alpha \alpha}(u) B_a(v) \ket 0= F^a \alpha_\alpha (u-v) B_\gamma(v) r_{\alpha a}^{\tau \gamma} (v-u) T_{k\tau}(u) \ket{0}+\dots=\\
& F^a  \alpha_2 (u-v) (-e^{2 i \phi})^{2-L} B_\gamma(v) \prod_{i=1}^L f(u,\{u_i\})  (-e^{2 i \phi})^{\alpha(L-1)}r_{\alpha a}^{\alpha \gamma} (v-u)\ket{0}+\dots,
\end{align}
where in the last line we used $\alpha_k=(-e^{2\,i\,\phi})^{2-k}\alpha_2$ from \eqref{condalpha}.\\
By neglecting the $\dots$ terms for the moment, we see that $\ket{1}$ is an eigenstate of the transfer matrix if
\begin{equation}
F^a \sum_{\alpha=1}^2 (-e^{2 i \phi})^{\alpha(L-1)}r_{\alpha a}^{\alpha \gamma} (v-u) \sim F^\gamma
\end{equation}
and expanding the sum we get
\begin{equation}
F^a (-e^{2 i \phi})^{L-1} (r_{1 a}^{1 \gamma} +(-e^{2 i \phi})^{L-1}r_{2 a}^{2 \gamma}) \sim F^\gamma,
\label{combr}
\end{equation}
so $F^a$ needs to be an eigenvector of the combination of $r$ given in \eqref{combr}. This will be more clear in the case of $M$ particles, but the contractions of the indices in the $r$ define the transfer matrix of the 6-vertex model for a spin chain of length 1. To summarize, if $F$ is an eigenstate of this transfer matrix, $T_{\alpha \alpha}$ acts diagonally on $\ket 1$. The initial problem of finding the eigenvalues of the transfer matrix built from the $R$-matrix of our model, reduces to the auxiliary problem to diagonalize the transfer matrix of 6-vertex type and \textit{here comes the nesting}. This last problem will be solved in the appendix \ref{betheansatznested}.\\
During all the discussion, we ignored the terms $\dots$. The condition for which those terms cancel is called \textit{Bethe equation} and fixes the value of the rapidity $v$. For the case of one particle, this calculation is still doable, but becomes very tedious for the states of more magnons. We followed here the standard shortcut that gives the same Bethe equations as the explicit calculation. The eigenvalue of the transfer matrix obtained by summing \eqref{T00one}, \eqref{T33one} and \eqref{Taaone} should be regular, the residue  at the pole should vanish. In this case the eigenvalues of the transfer matrix has two poles, for $u\to v$ and $u\to v+\psi$. In what follows we will require the cancellation of the residue at $u \to v$, but it can be proved that analysing the residue around the second pole will give a set of equation that can be mapped to the ones we are giving.\\
This leads to the following results
\begin{equation}
T(u)B_a \ket{0}=\Lambda_{1,a}(u)B_a\ket{0},
\end{equation}
where $N=1$ if the particle is created by $B_2$ and $N=0$ otherwise,
\begin{align}
\frac{\Lambda_{1,a}(u)}{\sigma e^{\psi +i \sigma \phi }}=&\frac{i \cosh \left(u-v+\psi \right)}{\sinh \left(v-u\right)}+\frac{i \cosh \left(u-v-2 \psi \right)}{\sinh \left(u-v-\psi \right)}\prod _{i=1}^L \frac{ \tanh \left(u-u_i\right) \sinh \left(u-u_i+\psi \right)}{e^{2 \psi }\cosh \left(u-u_i-\psi \right)}+\\
& i \left(-e^{2 i \phi }\right)^{N-L+1} \frac{\cosh \left(u-v-\psi \right)}{\sinh \left(u-v\right)} \lambda_{6V}(u) \prod _{j=1}^L \frac{ \sinh \left(-u_j+u+\psi \right)}{i e^{\psi +i \phi }\cosh \left(u-u_j\right)},
\end{align}
where $\sigma=1$ for $a=1$ ($N=0$) and $\sigma=-1$ for $a=2$ ($N=1$).\\
The expression of $\lambda_{6V}(u)$ will be derived in the appendix \ref{betheansatznested}, for completeness we will also write here the one for one particle
\begin{align}
\frac{\lambda _{6V}(u)}{\left(-e^{2 i \phi }\right)^{L-N-1}}=&\left(-e^{2 i \phi }\right)^L\left(\frac{\sinh \left(2 \left(u-w-\psi \right)\right)}{\sinh \left(2 \left(w-u\right)\right)}\right)^N \frac{\sinh \left(2 \left(v-u\right)\right)}{\sinh \left(2 \left(u-v-\psi \right)\right)}+\\
&\left(\frac{\sinh \left(2 \left(w-u-\psi \right)\right)}{\sinh \left(2 \left(u-w\right)\right)}\right)^N.
\end{align}
In order to cancel the unwanted term, the rapidities should satisfy the following condition
\begin{align}
\left(\frac{\sinh \left(2 \left(w-v-\psi \right)\right)}{\sinh \left(2 \left(v-w\right)\right)}\right)^N \prod _{j=1}^L \frac{ \sinh \left(v-u_j+\psi \right)}{i\, e^{i \phi +\psi }\cosh \left(v-u_j\right)}=1
\label{BEonep}
\end{align}
and
\begin{align}
\frac{\left(-e^{2 i \phi }\right)^L \sinh \left(2 \left(v-w\right)\right)}{\sinh \left(2 \left(w-v-\psi \right)\right)}=1.
\label{beonepnest}
\end{align}
We would like to clarify the meaning of all the parameters appearing in the expression. The $u_i$ are the set of inhomogeneities of the main chain. $v$ is the rapidity of the one magnon state we are considering and satisfy the Bethe equation \eqref{BEonep}. $v$ is also the inhomogeneity in the nested chain. If the magnon is created by $B_2$, $N=1$,  there is also the parameter $w$. This latter is the rapidity of the particle in the nested chain and can be calculated via the auxiliary Bethe equation \eqref{beonepnest}.\\
The block structures of the eigenvalue and the Bethe equations suggest a way to generalize this result to $M$ magnons. Furthermore, one can get the explicit expression of the eigenstate recursively by following the derivation given in \cite{martins1998quantum}.\\
To give a hint on how this method works and where the difficulties emerge for the $M$ magnons state, in the appendix \ref{2partstates} we will explicitly derive the expression for 2 magnons.

\subsection{M-particles state}
\label{Mparticlestates}
As already stressed, the eigenvalue of the transfer matrix for the $M$ particle states can be derived by generalizing the expressions for one and 2 magnons respectively in section \ref{onepstate} and in the appendix \ref{2partstates}. The expression of the eigenstate for the $M$ particle will involve a combinatorial expression due to the fact that $B_1$ and $B_2$ generates particles, but $B_3$ generates a pair.\\
However, we will now show that to find the expression of the eigenvalue and the Bethe equation we do not need to know the $M$ particle state explicitly.\\
Let us consider more closely the eigenvalues for the case of two particles derived in the appendix \ref{2partstates} in terms of the functions $\theta_a$, $\alpha_a$ and $\zeta_{1,a}$ defined in \eqref{alpha}-\eqref{zeta},
\begin{align}
\nonumber
\Lambda_2(u)=&\theta_{a_1}(u-v_1)\theta_{a_2}(u-v_2)+\\
\nonumber
&\zeta_{1,{a_1}}(u-v_1)\zeta_{1,{a_2}}(u-v_2)\prod_{j=1}^L \frac{\sinh \left(u-u_j+\psi \right)}{\cosh \left(u-u_j-\psi \right)} \frac{\tanh \left(u-u_j\right)}{e^{2 \psi }}+\\
&\left(-e^{2 i \phi }\right)^{4-L}\alpha_2(u-v_1)\alpha_2(u-v_2)\lambda_{6V}(u)\prod_{j=1}^L\frac{ \sinh \left(u-u_j+\psi \right)}{i\,e^{\psi +i \phi } \,\cosh \left(u-u_j\right)}.
\end{align}
The meaning of this eigenvalue is clear:
\begin{itemize}
\item the terms with $\theta$, $\zeta$ are the coefficients in the commutation relations $T_{00}$ and $T_{33}$ with each $B_a$,
\item the terms $\alpha$ and $\lambda_{6v}$ are in the commutator $T_{11}+T_{22}$ with the $B_a$,
\item the terms with the product $\prod_{i=1}^L$ comes from the action of $T_{ii}$ on the vacuum.
\end{itemize}
The eigenvalues appears as factorized products of single-excitations terms, so this strongly suggest that, even if the exact eigenstate of two particle state is \eqref{2particlesstate}, the eigenvalues can be obtained very naively just considering
\begin{align}
\ket{2\{v\}}\sim F^{ab}B_a(v_1)B_b(v_2)\ket{0}.
\end{align}
With this in mind, we generalize the result to arbitrary number $M$ of magnons. To do this, we act with the transfer matrix on the state
\begin{align}
\ket{M\{v\}}\sim F^{a_1 a_2 \dots a_M}B_{a_1}(v_1)\dots B_{a_M}(v_M)\ket{0},
\end{align}
where $N$ magnons are generated by $B_2$ and we get the following eigenvalue
\begin{align}
\nonumber\frac{\lambda(u)}{\mathcal{N}}=&\prod _{i=1}^M \frac{ \cosh \left(u-v_i+\psi \right)}{\sinh \left(v_i-u\right)}+\\
\nonumber&\frac{\lambda_{6V}(u)}{\left(-e^{2 i \phi }\right)^{L-M-N}} \prod _{i=1}^M  \frac{\cosh \left(u-v_i-\psi \right)}{\sinh \left(u-v_i\right)}\prod _{j=1}^L \frac{  \sinh \left(u-u_j+\psi \right)}{i\,e^{i \phi +\psi }\,\cosh \left(u-u_j\right)}+\\
&\prod _{i=1}^M \frac{ \cosh \left(u-v_i-2 \psi \right)}{\sinh \left(u-v_i-\psi \right)}\prod _{i=1}^L \frac{ \tanh \left(u-u_i\right) \sinh \left(u-u_i+\psi \right)}{e^{2 \psi }\,\cosh \left(u-u_i-\psi \right)},
\end{align}
%\comC{add shift}\\
where $\mathcal{N}=i^M \left(-e^{2 i \phi }\right)^{M-N} \left(-e^{\psi -i \phi }\right)^M$ and $\lambda_{6V}(u)$ is the eigenvalue of the auxiliary problem given in \eqref{eigenv6vexplicit}. For completeness we will also report it here
\begin{align}
\nonumber\frac{\lambda _{6V}(u)}{\left(-e^{2 i \phi }\right)^{L-M}}=&\prod _{i=1}^N \frac{ \sinh \left(2 \left(u-w_i+\psi \right)\right)}{e^{2 i \phi }\sinh \left(2 \left(u-w_i\right)\right)}+\\
&\left(-e^{2 i \phi }\right)^{L-M} \prod _{j=1}^N \frac{\sinh \left(2 \left(u-w_j-\psi \right)\right)}{e^{2 i \phi } \sinh \left(2 \left(u-w_j\right)\right)}\prod _{j=1}^M \frac{e^{2 i \phi } \sinh \left(2 \left(u-v_j\right)\right)}{\sinh \left(2 \left(u-v_j-\psi \right)\right)}.
\end{align}
%\textcolor{blue}{shift affed later}
%\\
To find the Bethe equation, we will use the same shortcut of the one particle case. We will impose that the eigenvalue of the transfer matrix is regular, so that the spurious pole  cancels. We derived the Bethe equation by requiring that the residue at the pole $u=v$ cancels. Another set of Bethe equations can be derived from $u=v+\psi$, but those are not independent to the ones found here, but can be mapped to them.\\
We found that the rapidities $\{v\}$ of the main chain should satisfy the constraint
\begin{align}
&\prod _{i=1,i\neq j}^M -\frac{\cosh \left(v_j-v_i+\psi \right)}{\cosh \left(v_j-v_i-\psi \right)}=\prod _{i=1}^L \frac{  \sinh \left(v_j-u_i+\psi \right)}{i\, e^{\psi +i \phi } \cosh \left(v_j-u_i\right)} \prod _{i=1}^N \frac{\sinh \left(2 \left(w_i-v_j-\psi \right)\right)}{\sinh \left(2 \left(v_j-w_i\right)\right)}
\label{BEmainchiain}
\end{align}
for $j=1,\dots, M$, while the $w$'s satisfy the auxiliary Bethe equations \eqref{BEnestedchain}, 
\begin{equation}
\prod _{i=1,i\neq j}^N \frac{\sinh \left(2 \left(w_j-w_i+\psi \right)\right)}{\sinh \left(2 \left(w_j-w_i-\psi \right)\right)}=\left(-e^{2 i \phi }\right)^{L}\prod _{k=1}^M \frac{\sinh \left(2 \left(w_j-v_k\right)\right)}{\sinh \left(2 \left(v_k-w_j+\psi \right)\right)},
\label{BEnestedchain}
\end{equation}
for $j=1,\dots,N$.

\subsection{General result}
The Bethe equations that we found before take a bit of a unusual form due to the presence of both $\cosh$ and $\sinh$. However, we can rewrite both the Bethe equations and the eigenvalue by considering a shift in $u_i$, $v_i$ and $\psi$,
\begin{align}
&\psi \to  {\Psi} +\frac{i \pi }{2},
&&u_i\to  \mathtt{u}_i+\frac{i \pi }{2},
&&v_i\to  \mathtt v_i-\frac{\psi }{2},
\end{align}
%\comC{shall we call them tilde?}
We remark, as mentioned in 2.1, that the coupling constant of the theory is $\gamma \to \tanh \psi$ while under this shift $\gamma\to \coth \Psi $. This means that under this map the strong and weak coupling regimes are interchanged. The eigenvalue of the transfer matrix now becomes
\begin{align}
\nonumber\frac{\lambda(u)}{\mathcal{N}}=&\prod _{i=1}^M \frac{\sinh \left(u-\mathtt v_i+\frac{3 \Psi }{2}\right)}{\sinh\left(u-\mathtt v_i+\frac{\Psi }{2}\right)}+\\
\nonumber&\frac{\lambda_{6V}(u)}{\left(-e^{2 i \phi }\right)^{L-M-N}} \prod _{i=1}^M  \frac{\sinh \left(u-\mathtt v_i-\frac{\Psi }{2}\right)}{\sinh\left(u-\mathtt v_i+\frac{\Psi }{2}\right)} \prod _{j=1}^L \frac{  \sinh \left(u-\mathtt u_j+\psi \right)}{i\,e^{\Psi +i \phi }\sinh\left(u-\mathtt u_j\right)}+\\
&\prod _{i=1}^M \frac{\cosh \left(u-\mathtt v_i-\frac{3 \Psi }{2}\right)}{\cosh \left(u-\mathtt v_i-\frac{\Psi }{2}\right)}\prod _{i=1}^L \frac{ \coth \left(u-\mathtt u_i\right) \sinh \left(u-\mathtt u_i+\Psi \right)}{e^{2 \Psi }\,\cosh \left(u-\mathtt u_i-\Psi \right)},
\end{align}
with $\mathcal{N}=\left(-e^{2 i \phi }\right)^{M-N} \left(-e^{\Psi -i \phi }\right)^M$ and for the nested chain
\begin{align}
\nonumber\frac{\lambda _{6V}(u)}{\left(-e^{2 i \phi }\right)^{L-M}}=&\prod _{i=1}^M-\frac{ \sinh \left(2 \left(u-w_i+\Psi \right)\right)}{e^{2 i \phi }\,\sinh\left(2 \left(u-w_i\right)\right)}+\\
&\left(-e^{2 i \phi }\right)^{L-M} \prod _{i=1}^N -\frac{ \sinh \left(2 \left(u-w_i-\Psi \right)\right)}{e^{2 i \phi } \sinh\left(2 \left(u-w_i\right)\right)}\prod _{i=1}^M -\frac{e^{2 i \phi } \sinh \left(2 u-2 \mathtt v_i+\Psi \right)}{\sinh\left(2 u-2 \mathtt v_i-\Psi \right)}.
\end{align}
Under the same shift, the Bethe equations become
\begin{align}
&\prod _{i=1,i\neq j}^M \frac{\sinh\left(\mathtt v_i-\mathtt v_j-\Psi \right)}{\sinh\left(\mathtt v_i-\mathtt v_j+\Psi \right)}=\prod _{i=1}^L \frac{1}{i\,e^{\Psi +i \phi }}\frac{  \sinh\left(\mathtt u_i-\mathtt v_j-\frac{\Psi }{2}\right)}{\sinh \left(\mathtt u_i-\mathtt v_j+\frac{\Psi }{2}\right)} \prod _{i=1}^N \frac{\sinh \left(2 \mathtt v_j-2 w_i+\Psi \right)}{\sinh\left(2 \mathtt v_j-2 w_i-\Psi \right)}
\end{align}
for $j=1,\dots,M$ and for the nested chain
\begin{equation}
\prod _{i=1,i\neq j}^N \frac{\sinh \left(2 \left(w_i-w_j-\Psi \right)\right)}{\sinh\left(2 \left(w_i-w_j+\Psi \right)\right)}=\left(-e^{2 i \phi }\right)^{L}\prod _{i=1}^M \frac{\sinh\left(2 \mathtt v_i-2 w_j-\Psi \right)}{\sinh\left(2 \mathtt v_i-2 w_j+\Psi \right)},
\end{equation}
for $j=1,\dots,N.$
%\textcolor{blue}{shift added for nested BE and BE}
\\
Let us introduce the standard Baxter Q-functions %\comC{here all u are the shifted ones, shall we use the same character ?}
\begin{align}
\label{Qbaxt1}&Q^{[a]}_{\mathtt u}(x) = \prod_{i=1}^L \sinh[x-{\mathtt u}_i-a\Psi] , &&\tilde{Q}^{[a]}_{\mathtt u}(x) = \prod_{i=1}^L \cosh[x-{\mathtt u}_i-a\Psi], \\
&Q^{[a]}_{\mathtt v}(x) = \prod_{i=1}^M \sinh[x-{\mathtt v}_i-a\Psi], &&\tilde{Q}^{[a]}_{\mathtt v}(x) = \prod_{i=1}^M \cosh[x-{\mathtt v}_i-a\Psi], \\
\label{Qbaxt2}&Q^{[a]}_w(x) = \prod_{i=1}^N \sinh[x-w_i-a\Psi], && \tilde Q^{[a]}_w(x) = \prod_{i=1}^N \cosh[x-w_i-a\Psi].
\end{align}
The eigenvalue is
\begin{align}
\frac{\lambda(u)}{\mathcal{N}}=&\frac{{Q_\mathtt{v}}^{[-3/2]}}{{Q_\mathtt{v}}^{[-1/2]}}+e^{-2\Psi L}\frac{{\tilde Q_\mathtt{v}}^{[3/2]}}{{\tilde Q_\mathtt{v}}^{[1/2]}} \frac{{\tilde Q_\mathtt{u}}^{[0]}}{{ Q_\mathtt{u}}^{[0]}}\frac{{Q_\mathtt{u}}^{[-1]}}{{\tilde Q_\mathtt{u}}^{[1]}}+\frac{ \left(-i e^{-\Psi -i \phi }\right)^L}{\left(-e^{2 i \phi }\right)^{L-M-N}} \frac{{Q_\mathtt{v}}^{[1/2]}}{{Q_\mathtt{v}}^{[-1/2]}}\frac{{Q_\mathtt{u}}^{[-1]}}{{Q_\mathtt{u}}^{[0]}}\lambda_{6V}(u),
\end{align}
\begin{align}
\lambda _{6V}(u)\frac{e^{2 i \phi N}}{\left(-e^{2 i \phi }\right)^{L-M}}=\frac{Q^{[-1]}_w\tilde{Q}^{[-1]}_w}{Q^{[0]}_w\tilde{Q}^{[0]}_w}+\left(-e^{2 i \phi }\right)^{L} \frac{Q^{[1]}_w\tilde{Q}^{[1]}_w}{Q^{[0]}_w\tilde{Q}^{[0]}_w}\frac{Q^{[-1/2]}_\mathtt{v}\tilde{Q}^{[-1/2]}_\mathtt{v}}{Q^{[1/2]}_\mathtt{v}\tilde{Q}^{[1/2]}_\mathtt{v}}
\end{align}\\
where for simplicity we used $Q^{[a]}_t(u)=Q^{[a]}_t$.\\
The Bethe equations for the main chain is
\begin{align}
\frac{{Q_\mathtt{v}}^{[-1]}}{{Q_\mathtt{v}}^{[1]}}=-\left(\frac{-i}{e^{\Psi+i \phi }}\right)^L \frac{{Q_\mathtt{u}}^{[-1/2]}}{{Q_\mathtt{u}}^{[1/2]}}\frac{{Q_w}^{[-1/2]}}{{Q_w}^{[1/2]}}\frac{{\tilde Q_w}^{[-1/2]}}{{\tilde Q_w}^{[1/2]}},
\end{align}
$j=1,\dots,M$ and for the nested chain
\begin{equation}
\frac{{Q_w}^{[-1]}{\tilde Q_w}^{[-1]}}{{Q_w}^{[1]} {\tilde Q_w}^{[1]}}=-\left(-e^{2 i \phi }\right)^L\frac{{Q_\mathtt{v}}^{[-1/2]} {\tilde Q_\mathtt{v}}^{[-1/2]}}{{Q_\mathtt{v}}^{[1/2]}{\tilde Q_\mathtt{v}}^{[1/2]}}.
\label{beqopnest}
\end{equation}
where  $Q^{[a]}_t(w_j)=Q^{[a]}_t$ and $j=1,\dots,N$.

\section{Numerical analysis}
\label{numerics}
In this section, we numerically evaluate the eigenvalues of the Lindblad superoperator for different values of the coupling constant $\psi$, the phase $\phi$ and the length of the spin chain $L$. Furthermore, we evaluated the scaling of the Liouvillian gap, the eigenvalue with the biggest real part, with the dimension of the spin chain.
\subsection{Numerical diagonalization of the transfer matrix}
\begin{figure}[h!]
	\centering
	\includegraphics[height=9.5cm]{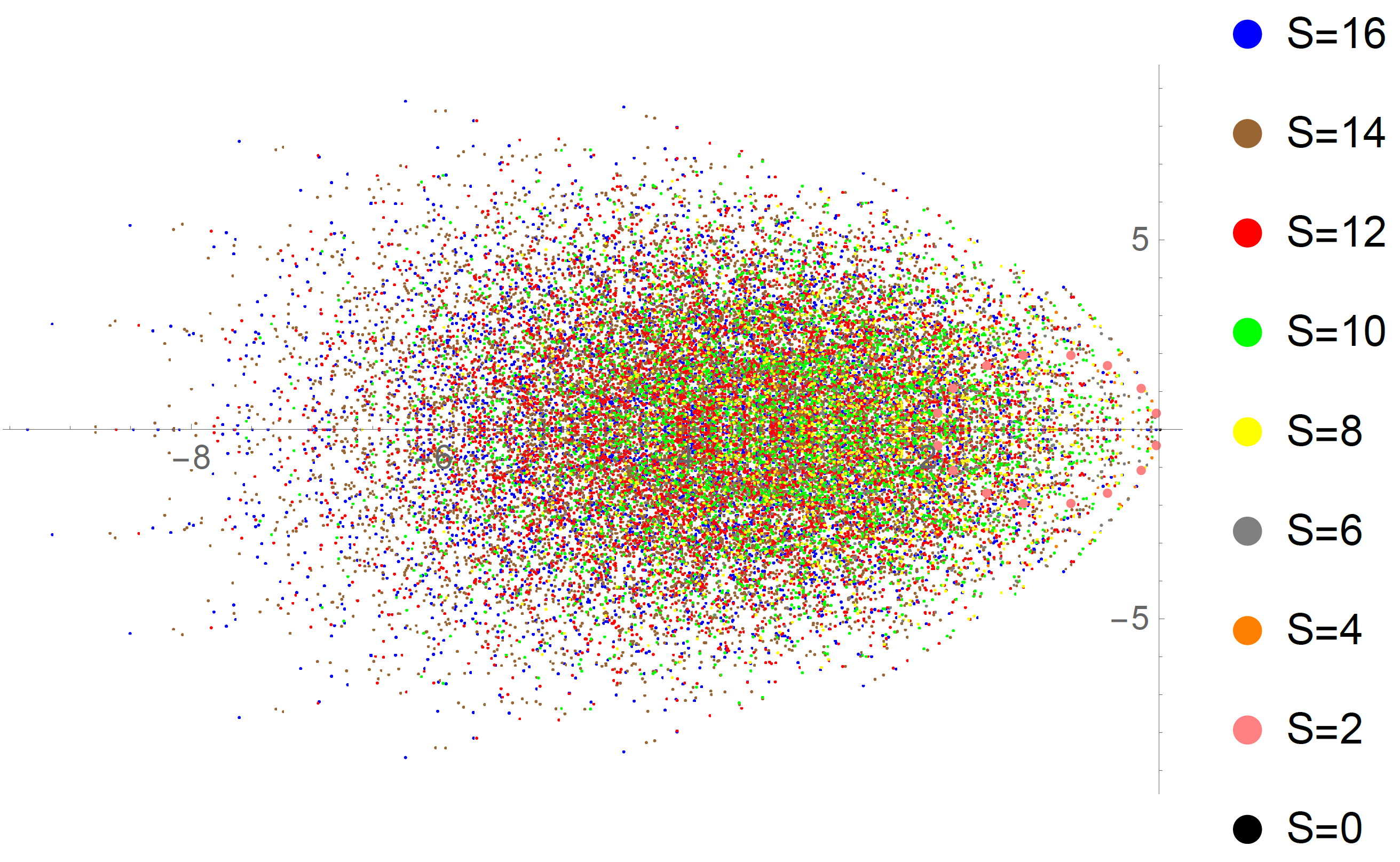}
	\caption{Numerical diagonalization of the transfer matrix for $L=8$ and $\psi=0.5$, $\phi=1$}	\label{psi05phi1}
	\includegraphics[height=9.5cm]{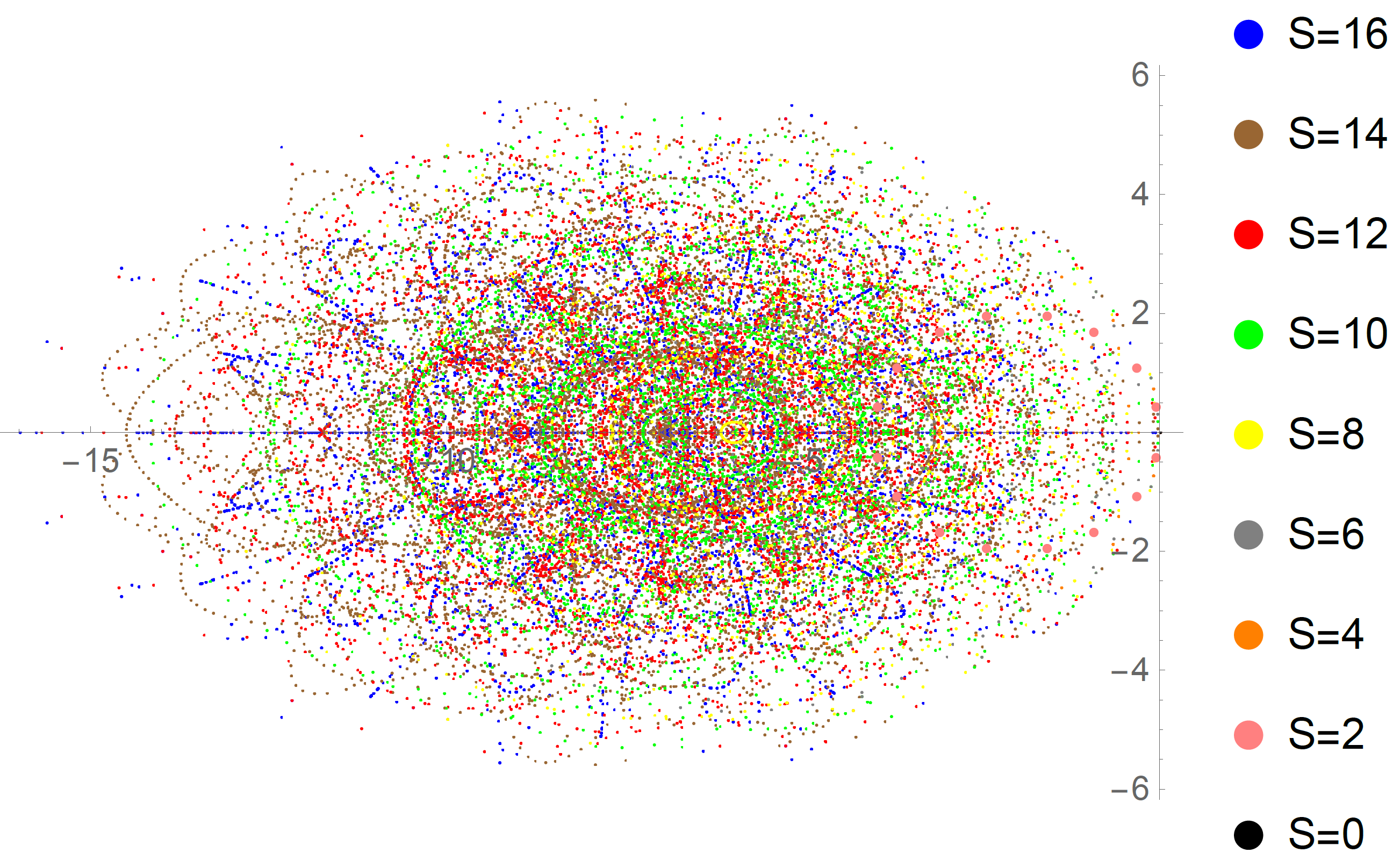}
	\caption{Numerical diagonalization of the transfer matrix for $L=8$ and $\psi=15$, $\phi=1$}	\label{psi15phi1}
	\end{figure}

In the numerical diagonalization of the transfer matrix, we considered three regimes of the coupling constants: $\psi=\frac{1}{100}$ (weakly coupling), $\psi=\frac{1}{2}$ (intermediate coupling) and $\psi=15$ (strongly coupling). The values of $\phi$ are also important, in fact it determine the nature of the non-equilibrium steady-state.\\
In open quantum systems, the non-equilibrium steady states are important objects of study. In the letter \cite{de2021constructing}, we found that the non-equilibrium steady states are mixed state. If the compatibility condition $e^{i (\phi+\pi/2)L}=1$ holds, it reduces to a pure state
\begin{align}
\rho_h=\ket{\Psi}\bra{\Psi},
\end{align}
where $\ket{\Psi}$ is a spin-helix state. We decided to study for each value of $\psi$, the value of $\phi$ characterizing a spin helix state, or a mixed state. Another point of interests is when the entropy of the state has a maximum, so it corresponds to a maximally mixed state. This point is reached when $\phi=\frac{2 \pi }{L}+ \frac{2 \pi c_1}{L}$, with $c_1$ integer.
\\
By numerical diagonalization, for the different values of $\psi$ and $\phi$, we found that for all the eigenvalues the real part is negative, similar behaviour as in \cite{essler-prosen-lindblad}. Furthermore, the behaviour for $\psi=\frac{1}{100}$ and $\psi=\frac{1}{2}$, was similar for all the values of $\phi$. We show the case $\psi=\frac{1}{2}$, $\phi=1$, Figure \ref{psi05phi1} and $\psi=15$, $\phi=1$, Figure \ref{psi15phi1}. \\
Our model conserves spin and hence we can decompose the Hilbert space into the different values of spin $S$. We notice that the sector with $S=2$, the eigenvalues lie on an ellipse. In the regime of strong coupling, the tori are preserved even in sector of higher spin. For the strong coupling regime, the tori structure is lost for the value of $\phi$ corresponding to a spin-helix state or a maximally mixed state.
\subsection{Evaluation of the Liouville gap}
\begin{figure}[h!]
	\centering
	\includegraphics[height=8cm]{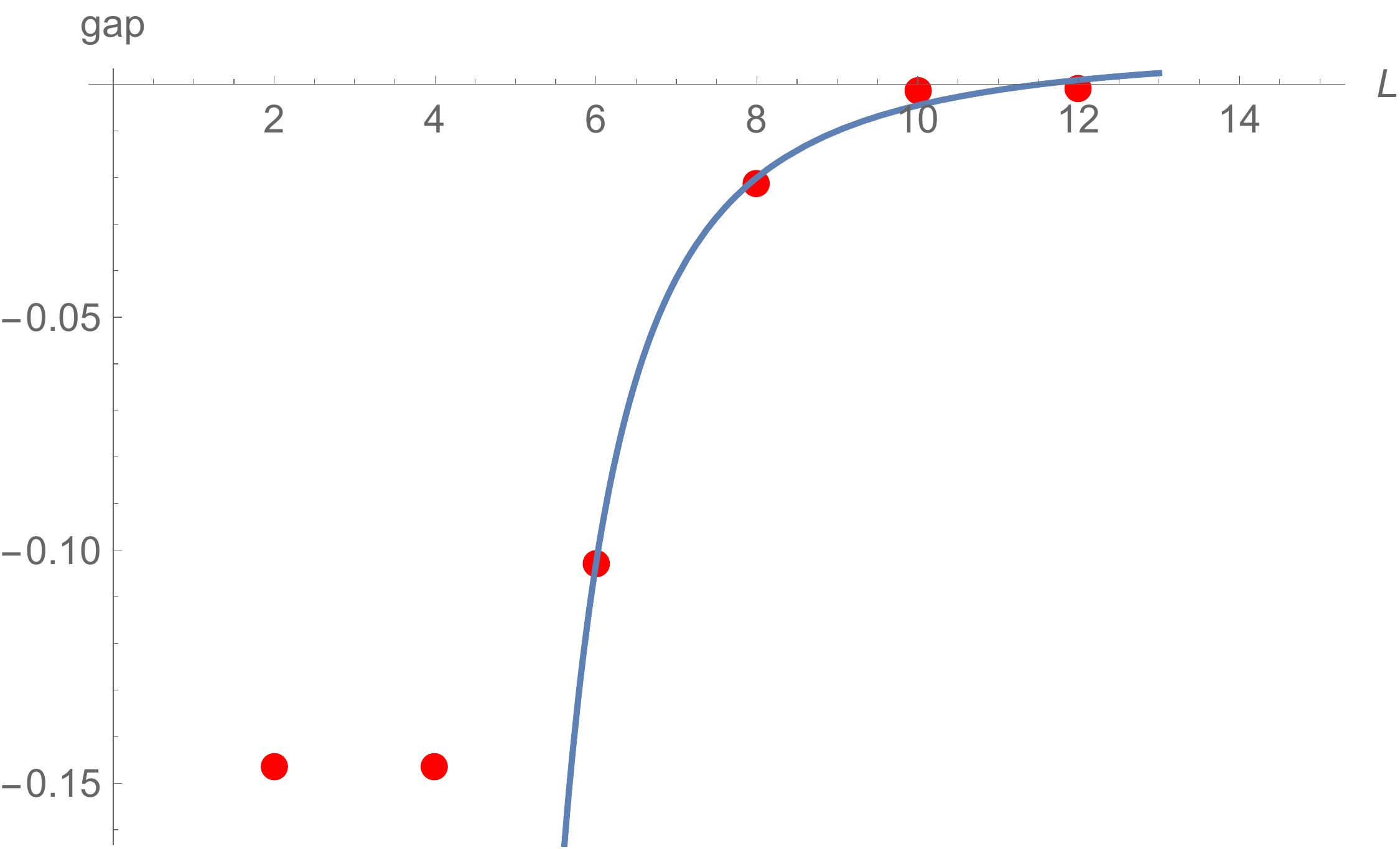}
	\caption{Gap for even length $L$ of the chain, $\psi=0.5$, $\phi=1.$ The red dots are the numerical calculation, the blue line is $\text{gap}\propto \frac{1}{L^2}$. The first two dots were not considered in the fix, they can be removed due to small size effect.}	\label{evenL}
	\includegraphics[height=8cm]{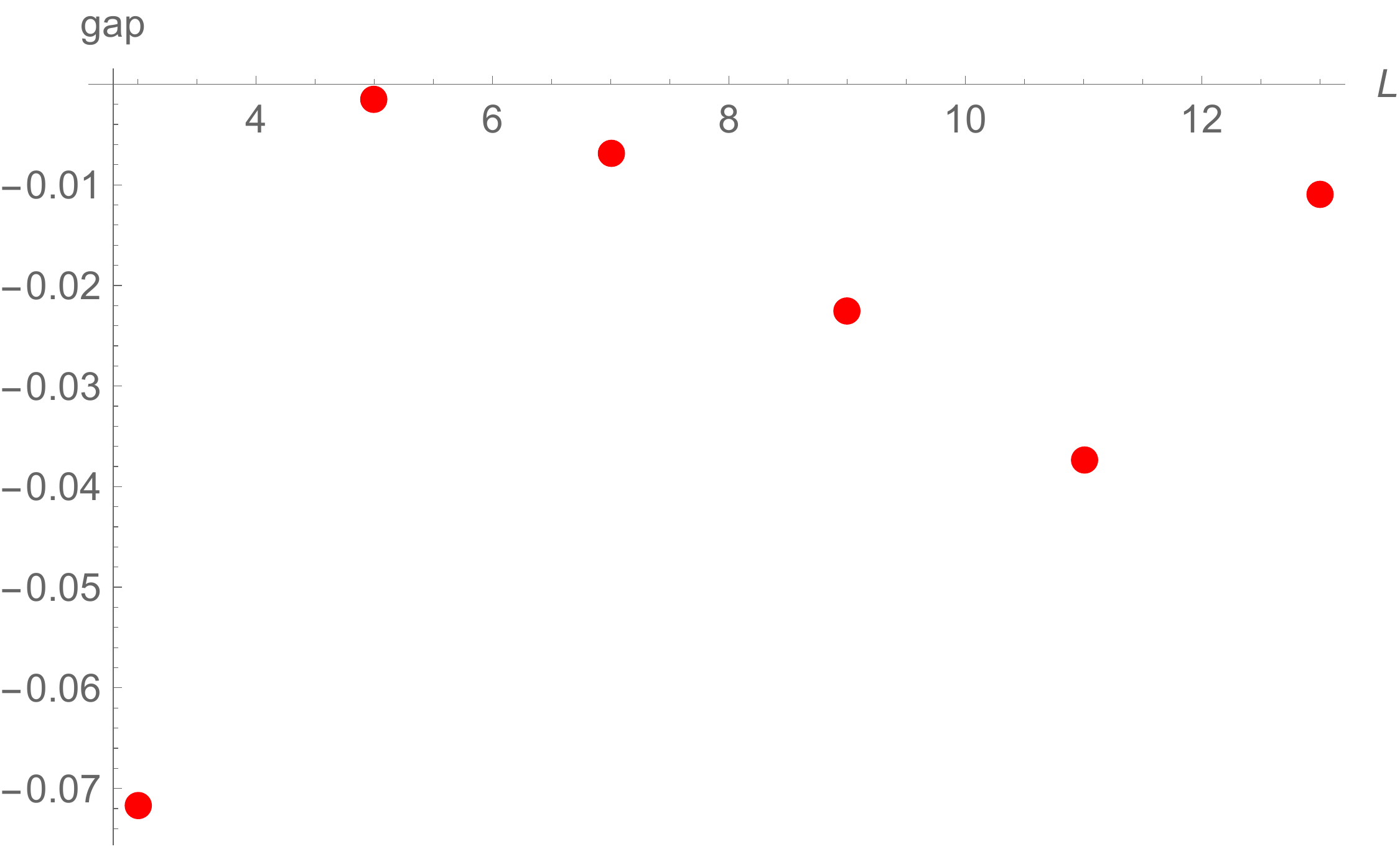}
	\caption{Gap for odd length $L$ of the chain, $\psi=0.5$, $\phi=1.$}	\label{oddL}
	\end{figure}

Subsequently we evaluated the gap, which corresponds to the eigenvalue with the second biggest real part (the smallest being 0). We analysed the dependence of the gap on the length of the spin chain. The behaviour for even or odd power of $L$ is different. In the first case, in fact, the gap scales as $1/L^2$, see Figure \ref{evenL}. The initial points in the graph were removed due to small size corrections. For $L$ odd, there is a bump corresponding to $L=11$, see Figure \ref{oddL}.

\begin{figure}[h!]
	\centering
	\includegraphics[height=8cm]{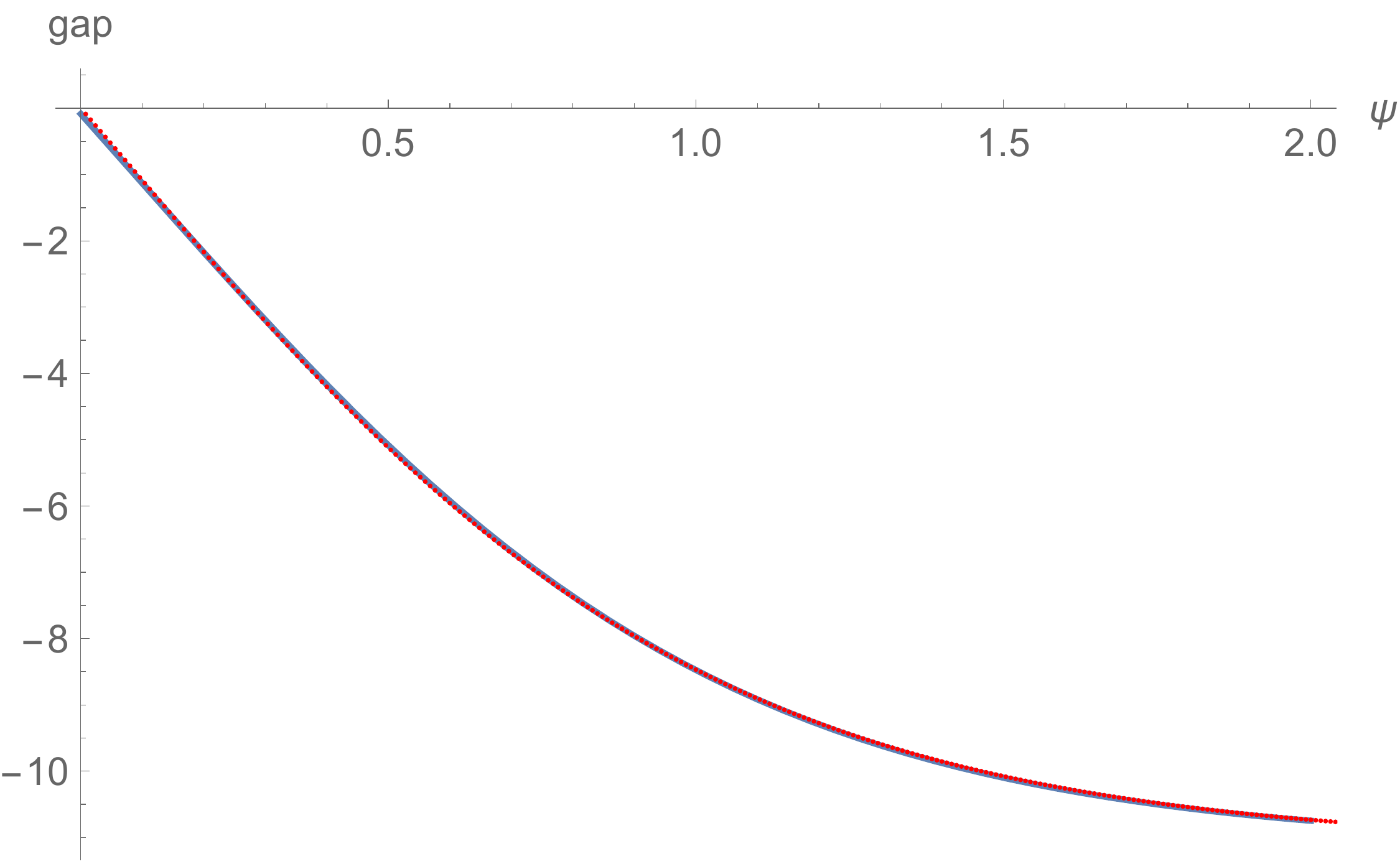}
	\caption{Gap for $L=8$ and $\phi=1.$ The red dots are the numerical calculation and the blue line is $\text{gap}\propto\tanh \psi$.}	\label{phidependence}
	\includegraphics[height=8cm]{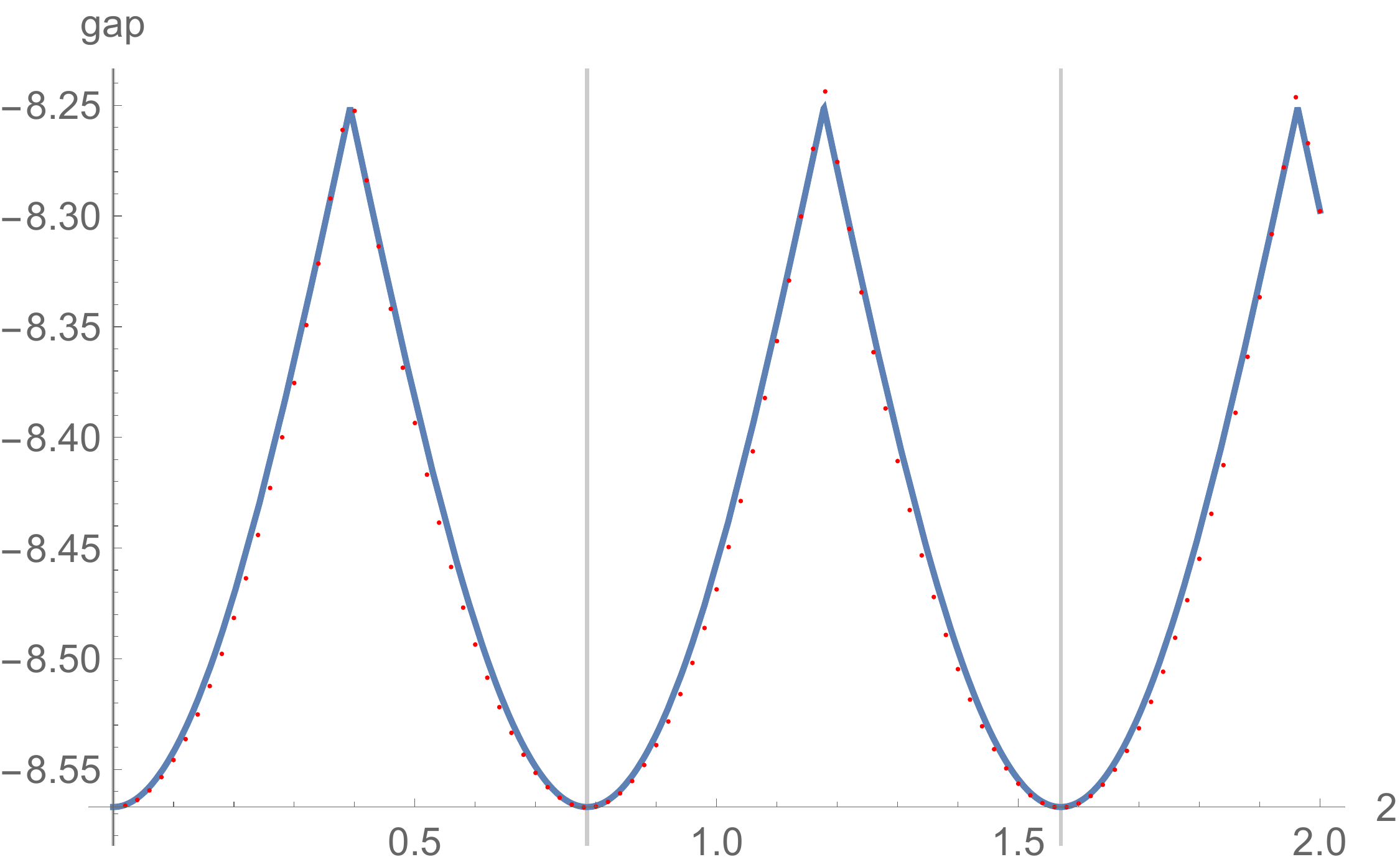}
	\caption{Gap for $L=8$ and $\psi=0.5.$ The red dots are the numerical calculation and the blue line is $\text{gap}\propto a+b |\cos \frac{\pi \phi}{c}|$.}	\label{psidependence}
	\end{figure}

With the computational power available, a brute force diagonalization of the gap was only feasible for small values of $L$. However, since model B3 preserves the spin we can work in a sector of given spin. A natural question to ask is: in which spin sector the states corresponding to the gap belong? By working on a state of given spin, the dimension of the matrix to analyse is smaller.\\
By considering the notation defined in section \ref{4dint}, we found that the eigenvalues of the operators
\begin{align}
&S^z=\sum_{i}\sigma^z_{i},
&&\tilde S^z=\sum_{i}\tilde \sigma^z_i,
&&S_{TOT}=S^z+\tilde S^z
\end{align}
are given in Table \ref{spin}.
\begin{table}[h!]
  \begin{center}
    \begin{tabular}{c|c|c|} 
      {L}  & {$S_z$} , {$\tilde S_z$} & {$S_{TOT}$}\\\hline
2&-2,2 & -2,2\\  \hline
3&-1,1 & 0\\  \hline
4&-4,-2,2,4 &-2,2 \\  \hline
5&-5,-3,3,5 & -2,2\\  \hline
6&-6,-4,4,6 & -2,2\\  \hline
7&-7,-5,5,7 & -2,2\\  \hline  
    \end{tabular}
  \end{center}
  \caption{Eigenvalues of the operators $S^z$, $\tilde S^z$, $S_{TOT}$.}
   \label{spin}
\end{table}
This suggest that the spin of the particle corresponding to the gap in a spin chain of length $L$ are $-L, -L+2, L-2, L+2$ and $S_{TOT}=-2, 2$. The case $L=3$ does not satisfy this due to small length effect. In this way, we obtained the gap until $L=13$, see Appendix \ref{trickforgap}.\\
We also evaluated the dependence of the gap on $\psi$. By fixing $\phi$ to different values, we found that the gap is proportional to $\tanh \psi$, as visible in Figure \ref{phidependence}. By evaluating the dependence on $\phi$, one notices that the gap is proportional to $a+b |\cos \frac{\pi \phi}{c}|$, see Figure \ref{psidependence}, where $c$ is a solution of the compatibility condition characterizing a spin helix state ($e^{i(\phi+\pi/2)L}=1$)  and $a$ and $b$ are constants depending on $L$.

\section{Discussion and conclusions}
In this paper we studied a new integrable open quantum system model found in \cite{de2021constructing}. It can be intepreted as two XXZ spin chains with an interaction term. We applied the nested algebraic Bethe ansatz to model B3. This allowed us to compute the analytical expression of the transfer matrix eigenvalue and consequently, by taking the logarithmic derivative, of all the conserved charges. To find the expression of the eigenvalue, we had to first solve the problem of diagonalizing a twisted transfer matrix of 6-V type. 
\\
By requiring that the residue at the pole of the eigenvalue cancel, we found the Bethe equation of the rapidities of the main chain and of the nested ones. We tested our results on spin chains of small length and up to 3 particles and found agreement.\\
In the second part of the paper, we numerically evaluated the  Liouville gap, which corresponds to the biggest real part of the eigenvalue of the transfer matrix. We found that for even $L$ it scales as $1/L^2$, where $L$ is the dimension of the system.\\
There are many interesting future directions in which this work could be extended. For example, one can try to recover the Liouville gap analytically from the expression of the eigenvalue. Another possibility would be to find an analogous of the string hypothesis and to the thermodynamic Bethe ansatz. Furthermore, the dynamics of model B3 seems to be highly non-trivial, the non-equilibrium states are mixed state and if they satisfy the compatibility conditions, they reduced to be spin helix state. Further work can be done in this direction, in particular by following \cite{popkov2017solution} one can analyse in which case of integrable open quantum system, the model admits spin-helix state.

\paragraph{Acknowledgements }   We would like to thank J. M. Nieto Garcìa, V. Popkov, B. Pozsgay, T. Prosen, A. L. Retore, A. Torrielli for useful discussions. We would like to thank R. Nepomechie for the code to check numerically the Bethe equations. We would like to thank A. L. Retore for valuable comments on the manuscript. MdL was supported by SFI, the Royal Society and the EPSRC for funding under grants UF160578, RGF$\backslash$R1$\backslash$181011, RGF$\backslash$EA$\backslash$180167 and 18/EPSRC/3590. 
C.P. is supported by the grant RGF$\backslash$R1$\backslash$181011

\appendix
\section{Bethe Ansatz for the nested chain}
\label{betheansatznested}
In this paper, via the Algebraic Bethe ansatz approach, we diagonalized the transfer matrix of an integrable open quantum system model. In this model, the nesting in manifest from the appearance of the transfer matrix for the twisted 6-vertex model. While the 6-vertex model also appears in the Hubbard model \cite{ramos1997algebraic,martins1998quantum} and in the $\text{AdS}_5 \times \text{S}_5$   $S$-matrix for bound states \cite{arutyunov2009bound}, the case analysed in the paper is different. For model B3, in fact, the $r$-matrix \eqref{rnested} is a twisted version of the standard one and the transfer matrix is also twisted as will be explained in the following.
\\
In principle, we can solve the nested problem as an independent one, for a spin chain of a given length and with arbitrary number of excitations. What we actually have to use is that the length of the chain is equal to $M$ (total number of magnons) and that the number of excitations of the nested chain is equal to $N$, number of excitation of type $B_2$. The rapidities of the particles of the main chain are the inhomogeneities of the nested chain.
\\
To summarize
\begin{align}
&M=\#B_1+\#B_2=L_{\text{nested}},
&&N=\#B_2=\#B,
\end{align}
where $B$ will be defined in the following as the creator operator in the nested chain.\\
The $r$-matrix of \eqref{rnested} is
\begin{align}
&r(u)=r_{\alpha \,\beta  }^{\gamma \,\delta}(u) e_\gamma^\alpha \otimes e_\delta^\beta,
&&r(u)=\left(
\begin{array}{cccc}
 1 & 0 & 0 & 0 \\
 0 & b(u)e^{-2 i \phi} & a(u) & 0 \\
 0 &  a(u) &b(u)e^{2 i \phi} & 0 \\
 0 & 0 & 0 & 1 \\
\end{array}
\right),
\end{align}
where $a(u)=\frac{\sinh (2 \psi )}{\sinh (2 (u+\psi ))}$, $b(u)=\frac{\sinh (2 u)}{\sinh (2 (u+\psi ))}$ and $\phi \in \mathbb{R}$.\\
As expected, if the twisting factor $e^{2 i \phi}=1$, one finds the standard 6-vertex $r$-matrix.\\
To construct the transfer matrix, we recall the results for one, two and three magnons
\begin{align}
&\text{one magnon:}\,&&(-e^{2 i \phi})^{\alpha(L-1)}r_{\alpha a}^{\alpha \gamma} (v-u)\\
&\text{two magnons:}\,&&(-e^{2 i \phi})^{\alpha(L-2)}r_{\alpha\, a_1}^{\tau\, \gamma} (v_1-u)r_{\tau\, a_2}^{\alpha\, \gamma_2} (v_2-u)\\
&\text{3 magnons:}\,&&(-e^{2 i \phi})^{\alpha(L-3)}r_{\alpha \,a_1}^{\tau_1 \gamma_1} (v_1-u)r_{\tau_1\, a_2}^{\tau_2 \, \gamma_2} (v_2-u)r_{\tau_2 \, a_3}^{\alpha \, \gamma_3} (v_3-u)
\end{align}
which can be easily generalized to the case of $M$ magnons
\begin{align}
(-e^{2 i \phi})^{\alpha(L-M)}r_{\alpha \,a_1}^{\tau_1 \gamma_1} (v_1-u)r_{\tau_1\, a_2}^{\tau_2 \, \gamma_2} (v_2-u)\dots r_{\tau_{M-1} \, a_M}^{\alpha \, \gamma_M} (v_M-u)
\label{Mmagnonstransfer}
\end{align}
and in matrix form
\begin{align}
&T^{(1)}(u)=\tr_a G_a \prod_{i=1}^M r_{a i}(u_i-u),\label{transferm}\\
&G_a=\left(-e^{2 i \phi }\right)^{\frac{3}{2} (M-L)} \left(
\begin{array}{cc}
 \left(-e^{2 i \phi }\right)^{\frac{M-L}{2}} & 0 \\
 0 & \left(-e^{2 i \phi }\right)^{\frac{L-M}{2}} \\
\end{array}
\right).
\end{align}
$T^{(1)}(u)$ is the twisted transfer matrix and $G_a$ is a $su(2)$ element. An example of twisted transfer matrix can be found in \cite{arutyunov2011twisting}. It is easy to check that the RTT is satisfied also with this new definition of transfer matrix.\\
In what follows, $^{(1)}$ identifies objects in the nested chain. Since the model preserves spin, the reference state is defined as a state with all spin up
\begin{equation}
\ket{0}^{(1)}=\bigotimes_{i=1}^M \ket{\uparrow}=\bigotimes_{i=1}^M \left(
\begin{array}{c}
 1 \\
 0 \\
\end{array}
\right).
\label{vacuumnested}
\end{equation}
The monodromy matrix is
\begin{equation}
T_a^{(1)}(u)=G_a \prod_{i=1}^M r_{ai}(v_i-u)=\left(
\begin{array}{cc}
 A(u) & B(u) \\
 C(u) & D(u) \\
\end{array}
\right),
\end{equation}
\begin{equation}
A(u)\ket{0}^{(1)}=\left(-e^{2 i \phi }\right)^{L-M} \ket{0}^{(1)},
\end{equation}
\begin{equation}
C(u)\ket{0}^{(1)}=0,
\end{equation}
\begin{equation}
D(u)\ket{0}^{(1)}=\left(-e^{2 i \phi }\right)^{2 (L-M)} \prod _{i=1}^M e^{2 i \phi }\, b\left(v_i-u\right)\ket{0}^{(1)}.
\end{equation}
It is easy to check that the reference state \eqref{vacuumnested} is an eigenstate of the transfer matrix with eigenvalue
\begin{align}
\left(-e^{2 i \phi }\right)^{L-M}\bigg[1+\left(-e^{2 i \phi }\right)^{L-M} \prod _{i=1}^M e^{2 i \phi }\, b\left(v_i-u\right) \bigg].
\end{align}
From the RTT relations\footnote{We can notice that the following commutation relations are the same of the ones for the untwisted transfer matrix, what changes is the action on the reference state.}
\begin{equation}
[B(u),B(v)]=0,
\end{equation}
\begin{equation}
A(v) B(u)= \frac{e^{-2 i \phi }}{b(v-u)}B(u) A(v)-e^{-2 i \phi } \frac{a(v-u) }{b(v-u)}B(v) A(u),
\end{equation}
\begin{equation}
D(v) B(u)= \frac{e^{-2 i \phi } }{b(u-v)}B(u) D(v)-e^{-2 i \phi }\frac{a(u-v)}{b(u-v)}B(v) D(u).
\end{equation}
The eigenstate of $N$ particles is given by
\begin{equation}
\ket{N\{w\}}^{(1)}=B\left(w_1\right)\dots B\left(w_N\right)\ket{0}^{(1)}.
\end{equation}
By applying the transfer matrix to a state of $N$ particles we found the following eigenvalue:
\begin{equation}
\frac{\lambda _{6V}(u)}{\left(-e^{2 i \phi }\right)^{L-M}}=\prod _{i=1}^N \frac{e^{-2 i \phi }}{b\left(u-w_i\right)}+\left(-e^{2 i \phi }\right)^{L-M} \prod _{i=1}^N \frac{e^{-2 i \phi }}{b\left(w_i-u\right)}\prod _{j=1}^M e^{2 i \phi } b\left(v_j-u\right)
\label{eigenv6v}
\end{equation}
and considering the expression for $a(u)$ and $b(u)$
\begin{align}
\nonumber\frac{\lambda _{6V}(u)}{\left(-e^{2 i \phi }\right)^{L-M}}=&\prod _{i=1}^N \frac{ \sinh \left(2 \left(u-w_i+\psi \right)\right)}{e^{2 i \phi }\sinh \left(2 \left(u-w_i\right)\right)}+\\
&\left(-e^{2 i \phi }\right)^{L-M} \prod _{i=1}^N \frac{\sinh \left(2 \left(u-w_i-\psi \right)\right)}{e^{2 i \phi } \sinh \left(2 \left(u-w_i\right)\right)}\prod _{j=1}^M \frac{e^{2 i \phi } \sinh \left(2 \left(u-v_i\right)\right)}{\sinh \left(2 \left(u-v_i-\psi \right)\right)}.
\label{eigenv6vexplicit}
\end{align}
This eigenvalue has a simple pole if $u=w_i$. We can require that the residue at the single pole vanishes and we get the set of Bethe equations for the rapidities $w$
\begin{equation}
\prod _{i=1, i\neq j}^N \frac{b\left(w_i-w_j\right)}{b\left(w_j-w_i\right)}=\left(-e^{2 i \phi }\right)^{L-M}\prod _{k=1}^M e^{2 i \phi } b\left(v_k-w_j\right),\,\,\,\,\, j=1,\dots,N
\end{equation}
or explicitly
\begin{equation}
\prod _{i=1,i\neq j}^N \frac{\sinh \left(2 \left(w_j-w_i+\psi \right)\right)}{\sinh \left(2 \left(w_j-w_i-\psi \right)\right)}=\left(-e^{2 i \phi }\right)^{L}\prod _{k=1}^M \frac{\sinh \left(2 \left(w_j-v_k\right)\right)}{\sinh \left(2 \left(v_k-w_j+\psi \right)\right)}.
\label{BEnestedchain}
\end{equation}

\section{Two particle state}
\label{2partstates}
In this appendix, we will explicitly compute the eigenvalues and the Bethe equations for a state of 2 magnons. The general steps here are slightly different than the one for the one-particle state. In fact, in this case, the state is composed by two parts
\begin{align}
\ket{2\{v\}}=B_{a_1}(v_1)B_{a_2}(v_2)F^{a_1 a_2}\ket{0}+B_3(v_1)g(v_1,v_2)\epsilon_{a_1 a_2}F^{a_1 a_2}T_{00}(v_2) \ket{0},
\label{2particlesstate}
\end{align}
the first one takes into account that the two particles are created by the operators $B_1$ and $B_2$. In a state of two particles there may also be a pair, created by $B_3$. The fermionic nature of the particle is manifest in the $\epsilon$ which considers the Pauli exclusion principle. The operator $T_{00}(v_2)$ in the second part is put for  dimensionality, in fact from \eqref{T00vacuum} the action of it on the vacuum is $1$.\\
In this appendix, we derive the expression of $g(u)$ and we interpret $F^{a_1 a_2}$ as the eigenvector on the transfer matrix in the nested chain.
\subsection{Action of $T_{00}$}
In order to evaluate the action of $T_{00}$ on $\ket {2,\{v\}}$ we need an additional commutation relation. In particular
\begin{align}
T_{00}(u)B_3(v)=q_1 B_3(v) T_{00}(u)+q_2 B_3(u) T_{00}(v)+q_3 B_{1}(u) B_{2}(v)+q_4 B_{2}(u) B_{1}(v),
\end{align}
where we omitted the dependence $\omega=\omega(u-v)$ on the coefficients.
\begin{align}
&q_1(u)=\frac{1}{2} e^{2 \psi } \text{csch}u\, (\cosh (2 u+3 \psi )+\cosh (\psi )) \text{csch}(u+\psi ),\\
&q_2(u)=-e^{u+2 \psi } \cosh \psi \, \text{csch}(u+\psi ) (\text{csch }u \cosh (u+2 \psi )-1),
\end{align}
\begin{align}
&q_3(u)=i \,e^{i \phi } \cosh\, \psi \, \text{csch}(u+\psi ),
&&q_4(u)=-i\, e^{-i \phi } \cosh \psi \, \text{csch}(u+\psi ).
\end{align}
By using this relation and \eqref{T00B}, we imposed  that
\begin{align}
T_{00}(u)\ket{2,\{v\}}=\lambda_0(u) \ket{2,\{v\}},
\end{align}
and uniquely fix the form of $g(v_1,v_2)$. We got
\begin{align}
g(v_1,v_2)=\frac{\theta_{a_1}(u-v_1)\rho_{a_2}(u-v_2)\eta(u-v_1)}{\lambda_0(u)-q_1(u-v_1)},
\end{align}
where $\eta(u)$ was defined in \eqref{eta}. $g(v_1,v_2)$ does not depend on $u$, in fact plugging all the expressions we got
\begin{align}
g(v_1,v_2)=i e^{2 \psi } \cosh \psi  \text{csch}\left(v_1-v_2-\psi \right)=\eta(v_1-v_2).
\label{expressiong}
\end{align}

This fix the expression of the ansatz for the 2 magnons state. The eigenvalue is
\begin{align}
\lambda_{0}(u)=\theta_{1,a_1}(u-v_1)\theta_{1,a_2}(u-v_2).
\label{eigv2p0}
\end{align}
This eigenvalue factorizes as a product of one particle eigenvalue. This is a strong hint on how to generalize the calculation to the case of $M$ magnons.
\subsection{Action of $T_{33}$}
Similarly, to confirm our result, we can act with $T_{33}$ on $\ket{2\{v\}}$. In this case, the commutation relations that we need are
\begin{align}
T_{33}(u)B_3(v)=\eta_1 B_3(v) T_{33}(u)+\eta_2 B_3(u) T_{33}(v)+\eta_3 T_{13}(u) T_{23}(v)+\eta_4 T_{23}(u) T_{13}(v),
\label{commrel1}
\end{align}
where we omitted the dependence $\omega=\omega(u-v)$ and
\begin{align}
&\eta_1(u)=\frac{1}{2} e^{2 \psi } \text{csch }u\, \text{csch}(u-\psi )(\cosh (2 u-3 \psi )+\cosh \psi ) ,\\
&\eta_2(u)=e^{u+2 \psi } \cosh \psi\,  \text{csch}(u-\psi ) (1-\text{csch }u \cosh (u-2 \psi )),
\end{align}
\begin{align}
&\eta_3(u)=i \cosh \psi\,  e^{2 \psi -i \phi } \text{csch}(u-\psi ),
&&\eta_4(u)=-e^{2 i \phi } \eta_3(u)
\end{align}
and
\begin{align}
T_{3a}(u)B_b(v)=\Gamma_{1ab} B_a(v) T_{3b}(u)+\Gamma_{2ab} B_b(v) T_{3a}(u)+\Gamma_{3} \epsilon_{ab}(B_3(v) C_3(u)-T_{00}(v) T_{33}(u)),
\label{commrel2}
\end{align}
\[ 
 \frac{2\,\Gamma_{1ab}(u) }{e^{2 \psi } \text{csch}u \,\text{csch}(u-\psi )}= 
  \begin{dcases*} 
\cosh (2 u-3 \psi )+\cosh \psi , & if  a=b=1,2 \\ 
\cosh \psi +\cosh (3 \psi ), & if  a$\neq$ b=1,2 
  \end{dcases*} 
\]

\[ 
\Gamma_{2ab}(u)= 
  \begin{dcases*} 
0, & if  a=b=1,2 \\ 
e^{2 \psi +2 i \phi }, & if  a=1, b=2\\ 
e^{2 \psi -2 i \phi }, & if  a=2, b=1
  \end{dcases*} 
\]
\begin{align}
\Gamma_3(u)=i\,e^{2 \psi } \cosh \psi  \, \text{csch}(u-\psi ).
\end{align}
By using \eqref{commrel1} and \eqref{commrel2} and the fact that $C_3(u)\ket{0}=T_{3\alpha}(u)\ket{0}=0$, imposing that
\begin{align}
T_{33}(u)\ket{2}=\lambda_3(u) \ket{2},
\end{align}
uniquely fix the form of $g(v_1,v_2)$. We got
\begin{align}
g(v_1,v_2)=\frac{-\zeta_2(u-v_1)\Gamma_3(u-v_2)}{\zeta_{1,a_1}(u-v_1)\zeta_{1,a_2}(u-v_2)-\eta_1(u-v_1)},
\end{align}
where $\eta(u)$ was defined in \eqref{eta}. Also in this case, $g(v_1,v_2)$ is independent on $u$ and coincides with \eqref{expressiong}.
\\
The eigenvalue of $T_{33}(u)$ is
\begin{align}
\lambda_{3}(u)=\zeta_{1,a_1}(u-v_1)\zeta_{1,a_2}(u-v_2)\prod_{i=1}^L e^{-2 \psi }\frac{\sinh \left(u-u_i+\psi \right)}{\cosh \left(u-u_i-\psi \right)} \tanh \left(u-u_i\right).
\label{eigv2p3}
\end{align}
We can notice that also in this case the eigenvalue factorizes as a product of one particle eigenvalue.
\subsection{Action of $T_{11}+T_{22}$}
This calculation is really important because it makes clear the appearance of the twisted transfer matrix.\\
The additional commutation relations that we need are
\begin{align}
C_{\alpha}(u)B_\beta (v)=\kappa_{\alpha \beta} e^{2 \psi}B_\beta(v)C_\alpha(u)+z_{1\alpha}T_{00}(v)T_{\alpha\beta}(u)+z_{2\alpha}T_{00}(u)T_{\alpha\beta}(v),
\end{align}
where the dependence of the $z$s on the spectral parameter is $z=z(u-v)$ and
\begin{align}
&\kappa_{11}=e^{2 i \phi }=\kappa_{22}^*,
&&\kappa_{12}=-1=-\kappa_{21},
\end{align}
\begin{align}
&\frac{i z_{11}(u)}{e^{i \phi }}=-i z_{12}(u) e^{i \phi }=i z_{22} (u) e^{i \phi }=-\frac{i z_{21}(u)}{e^{i \phi }}=\frac{\text{csch }u}{\tanh \psi-1}
\end{align}
and
\begin{align}
T_{\text{aa}}(u) B_3(v)= &s B_3(v) T_{11}(u)+s_2 \left(B_3(u) T_{\text{aa}}(v)-B_3(v) T_{\text{aa}}(u)\right)+\\
&s_{3,a} B_a(u) T_{\text{a3}}(v)+s_{4,a} T_{\text{a3}}(u) B_a(v)+s_5 B_3(v) T_{\text{aa}}(u),
\end{align}
where the dependence of the $s$s on the spectral parameter is $s=s(u-v)$ and
\begin{align}
&s= -(\tanh (\psi )-1)^{-2},
&&s_2(u)= \text{csch}^2\,u,
&&s_5=\tanh ^2\psi 
\end{align}
\begin{align}
\frac{s_{3,1}(u)}{i e^{i \phi }}=\frac{s_{4,1}(u)}{i e^{-i \phi }}=\frac{s_{3,2}(u)}{i e^{-i \phi }}=\frac{s_{4,2}(u)}{i e^{i \phi }}= (\tanh \psi -1) \text{ csch }u.
\end{align}
As in the previous cases, we need to identify
\begin{align}
\sum_{\alpha=1}^2 T_{\alpha\alpha}(u)\ket{2\{v\}}=\lambda_{12}(u)\ket{2\{v\}}.
\end{align}
After a very long calculation, one can separate the terms with two $B$s operators and the part with $B_3$.\\
From the second part, one can derive the expressions of $g(v_1,v_2)$ already derived in the previous two cases.\\
%\comC{I haven't done it, but should be the same. I tried but it is really long and i was making mistakes. It is really long because the term containing $B_3$ is not directly the one coming from the second part, but it mix with the first.\\
%Probably if the g from the previous two match, it should be fine even if we don't check it?}\\
From the first part, we get
\begin{align}
&\alpha_\alpha(u-v_1)\alpha_\alpha(u-v_2)r_{\alpha a_1}^{\tau \gamma}(v_1-u)r_{\tau a_2}^{\eta k}(v_2-u) B_\gamma(v_1) B_k (v_2) T_{\alpha\eta}(u)F^{a_1 a_2}\ket{0}\\
&\sim \lambda_{12}(u)F^{\gamma k}B_\gamma(v_1)B_k (v_2)\ket{0},
\end{align}
which can be simplified by considering that
$\alpha_k=(-e^{2 i \psi})^{2-k} \alpha_2$ and
\begin{align}
T_{\alpha\eta}(u)\ket 0=\delta_{\alpha \eta}\prod_{i=1}^L (-e^{2i \phi})^{\alpha-1}f(u)\ket{0},
\end{align}
and $f(u)=\frac{ \sinh \left(u-u_i+\psi \right)}{i\, e^{\psi +i \phi }\cosh \left(u-u_i\right)}$,
\begin{align}
&\alpha_2(u-v_1)\alpha_2(u-v_2)(-e^{2i \phi})^{2(2-\alpha)}r_{\alpha a_1}^{\tau \gamma}(v_1-u)r_{\tau a_2}^{\alpha k}(v_2-u) B_\gamma(v_1) B_k (v_2) \\&\prod_{i=1}^L (-e^{2i \phi})^{\alpha-1}f(u) F^{a_1 a_2}\ket{0}\sim \lambda_{12}(u)F^{\gamma k}B_\gamma(v_1)B_k (v_2)\ket{0},
\end{align}
similarly to the case for one particle, we found that the vector $F^{ab}$ should be an eigenvector of
\begin{align}
(-e^{2i \phi})^{2(2-\alpha)}r_{\alpha a_1}^{\tau \gamma}(v_1-u)r_{\tau a_2}^{\alpha k}(v_2-u)
\end{align}
which is the twisted transfer matrix of a chain of length 2. If this happens, the action of $\sum_{\alpha=1}^2 T_{\alpha\alpha}$  on $\ket{2 \{v\}}$ is diagonal. From this result, it is also straightforward to generalize the result for the case of $M$ magnons, as done in expression\footnote{For the expression in matrix form, see \eqref{transferm}.} \eqref{Mmagnonstransfer}.\\
We found that the eigenvalue of $T_{11}(u)+T_{22}(u)$ is
\begin{align}
\lambda_{12}(u)=\left(-e^{2 i \phi }\right)^{4-L}\alpha_2(u-v_1)\alpha_2(u-v_2)\lambda_{6V}(u)\prod_{i=1}^L\frac{  \sinh \left(u-u_j+\psi \right)}{i\,e^{\psi +i \phi }\,\cosh \left(u-u_j\right)}.
\label{eigv2p}
\end{align}
\subsection{Eigenvalue of the transfer matrix}
By summing the results \eqref{eigv2p0}, \eqref{eigv2p3}, \eqref{eigv2p}, we got the eigenvalue of the transfer matrix for the two magnon state and it is
\begin{align}
\nonumber
\Lambda_2(u)=&\theta_{a_1}(u-v_1)\theta_{a_2}(u-v_2)+\\
\nonumber
&\zeta_{1,{a_1}}(u-v_1)\zeta_{1,{a_2}}(u-v_2)\prod_{i=1}^L \frac{\sinh \left(u-u_i+\psi \right)}{\cosh \left(u-u_i-\psi \right)} \frac{\tanh \left(u-u_i\right)}{e^{2 \psi }}+\\
&\left(-e^{2 i \phi }\right)^{4-L}\alpha_2(u-v_1)\alpha_2(u-v_2)\lambda_{6V}(u)\prod_{i=1}^L\frac{ \sinh \left(u-u_j+\psi \right)}{i\,e^{\psi +i \phi } \,\cosh \left(u-u_j\right)}.
\end{align}
The structure of this eigenvalue appears in the form of factorized products of single-excitations terms. In section \ref{Mparticlestates}, we will start from it to find the general expression for the $M$ particle state and using it, by using the shortcut of the residue, we will find the Bethe equations.

\section{Shortcut to find the gap}
\label{trickforgap}
In this appendix, we give a shortcut to find the gap, the eigenvalue with biggest real part. As already mentioned, since model B3 preserves the spin, we can project the Hamiltonian through all the different spin sectors and diagonalize each of  the \textit{reduced} Hamiltonians.\\
From the table \ref{spin}, we found that in a spin chain of length $L$, the state producing the gap belong to the subsector characterized by $S^z=-L$. In what follows, we give a trick on how to select which of the vectors of the canonical basis are eigenvector in the sector of spin $S_z=-L$. In other words, we can consider the position of the entries $"-L"$ in the diagonal of $S_z$.\\
It is easy to check analytically that for a spin chain of length $L$, the $(2^L)^{th}$ entries of $S_z$ is $-L$. Starting from this entry and moving to the upper-left of the diagonal, for $L=2,3,4,5,$ it can be shown that the next $"-L"$ are in position $2^L-1$ and $2^L-4$. By calling $p_1=2^L$, $p_2=2^L-1$, $p_3=2^L-4$ and $p_4,\dots, p_{2^L}$ respectively the positions of all the other entries with $p_i<p_j$, for $i>j,$ we calculated the differences $\{p_{2^L}-p_{2^L-1},p_{2^L-1}-p_{2^L-2},\dots,p_{2}-p_{1}\}$, see Table \ref{tabledifferences}.

\begin{table}[h!]
  \begin{center}
    \begin{tabular}{c|c|} 
      {L}  & elements \\\hline
2&$\{1,3,1\},$ \\  \hline
3&$\{1,3,1,11,1,3,1\},$\\  \hline
4&$\{1,3,1,11,1,3,1,43,1,3,1,11,1,3,1\},$\\  \hline
5&$\{1,3,1,11,1,3,1,43,1,3,1,11,1,3,1,171,1,3,1,11,1,3,1,43,1,3,1,11,1,3,1\}$\\  \hline
    \end{tabular}
  \end{center}   
  \caption{Differences in position $p_i-p_{i-1}$ between the entries $"-L"$ of the diagonal of the operator $S_z$.}
 \label{tabledifferences}
\end{table}

We can notice the pattern
\begin{align}
\text{sequence}_L=\{\text{sequence}_{L-1}\,x_L\,\text{sequence}_{L-1}\}
\end{align}
where the number $x_L=\sum_{i=1}^{L-1} x_i =4^{L-1}$. \\Furthermore, the sequences of this number 
$1,3,11,43,171,683,2731,10923,43691$ is
\begin{equation}
a(n)=\frac{1}{3}(2^{2n+1}+1),\,\,\,\,\,n=0,1,\dots.
\end{equation}
By using this trick, we easily found\footnote{The limit of $L$ is given by the ram memory of the computer used.} the basis for the subspace of spin $s=-L$ until $L=13$. With this result, we compute the Hamiltonian in a sector of given spin
\begin{equation}
H_{red}=\sum_{ij} (v_i)^T \mathcal{L}_{i,j} v_j e_i^j,
\end{equation}
where $v_i$, element of the canonical basis is $(v_i)_j=\delta_{p_i,j}$.\\
By using this shortcut, we evaluated the gap until $L=13$, while with the direct diagonalization only until $L=8$.

\bibliography{BApaperbbl}

\end{document}